\documentclass[longauth]{aa}
\usepackage{pdflscape}
\usepackage{graphicx}
\usepackage{txfonts}
\usepackage{textcomp, gensymb}
\usepackage{lscape}
\usepackage[dvipsnames]{xcolor}
\usepackage{placeins}
\usepackage[version=4]{mhchem} 
\usepackage[colorlinks = true,
            linkcolor = blue,
            urlcolor  = blue,
            citecolor = blue,
            anchorcolor = blue, unicode]{hyperref}

\begin{document}

   \title{MINDS. JWST-MIRI reveals a peculiar \ce{CO2}-rich chemistry in the drift-dominated disk CX Tau}


   \author{Marissa Vlasblom\inst{1}
          \and
          Milou Temmink\inst{1}
          \and
          Sierra L. Grant\inst{2}
          \and
          Nicolas Kurtovic\inst{2}
          \and 
          Andrew D. Sellek\inst{1}
          \and
          Ewine F. van Dishoeck\inst{1,2}
          \and
          Manuel G\"udel \inst{3,4} 
          \and 
          Thomas Henning \inst{5}
          \and
          Pierre-Olivier Lagage \inst{6}
          \and
          David Barrado \inst{7}
          \and
          Alessio Caratti o Garatti \inst{8,9}
          \and
          Adrian M. Glauser \inst{4}
          \and
          Inga Kamp \inst{10}
          \and
          Fred Lahuis \inst{11}
          \and
          G\"oran Olofsson \inst{12}
          \and 
          Aditya M. Arabhavi \inst{10}
          \and
          Valentin Christiaens \inst{13,14}
          \and
          Danny Gasman \inst{13}
          \and
          Hyerin Jang \inst{15}
          \and
          Maria Morales-Calder\'on \inst{7}
          \and
          Giulia Perotti \inst{5}
          \and
          Kamber Schwarz \inst{5}
          \and
          Beno\^it Tabone \inst{16}}

   \institute{Leiden Observatory, Leiden University, 2300 RA Leiden, Netherlands\\ 
              \email{vlasblom@strw.leidenuniv.nl}
    \and
    Max-Planck Institut f\"{u}r Extraterrestrische Physik (MPE), Giessenbachstr. 1, 85748, Garching, Germany 
    \and
    Dept. of Astrophysics, University of Vienna, T\"urkenschanzstr. 17, A-1180 Vienna, Austria 
    \and
    ETH Z\"urich, Institute for Particle Physics and Astrophysics, Wolfgang-Pauli-Str. 27, 8093 Z\"urich, Switzerland 
    \and 
    Max-Planck-Institut f\"{u}r Astronomie (MPIA), K\"{o}nigstuhl 17, 69117 Heidelberg, Germany 
    \and
    Universit\'e Paris-Saclay, Universit\'e Paris Cit\'e, CEA, CNRS, AIM, F-91191 Gif-sur-Yvette, France 
    \and
    Centro de Astrobiolog\'ia (CAB), CSIC-INTA, ESAC Campus, Camino Bajo del Castillo s/n, 28692 Villanueva de la Ca\~nada, Madrid, Spain 
    \and
    INAF – Osservatorio Astronomico di Capodimonte, Salita Moiariello 16, 80131 Napoli, Italy 
    \and
    Dublin Institute for Advanced Studies, 31 Fitzwilliam Place, D02 XF86 Dublin, Ireland 
    \and
    Kapteyn Astronomical Institute, Rijksuniversiteit Groningen, Postbus 800, 9700AV Groningen, The Netherlands 
    \and
    SRON Netherlands Institure for Space Research, PO Box 800, 9700 AV, Groningen, The Netherlands 
    \and
    Department of Astronomy, Stockholm University, AlbaNova University Center, 10691 Stockholm, Sweden 
    \and
    Institute of Astronomy, KU Leuven, Celestijnenlaan 200D, 3001 Leuven, Belgium 
    \and
    STAR Institute, Universit\'e de Li\`ege, All\'ee du Six Ao\^ut 19c, 4000 Li\`ege, Belgium 
    \and
    Department of Astrophysics/IMAPP, Radboud University, PO Box 9010, 6500 GL Nijmegen, The Netherlands 
    \and 
    Universit\'e Paris-Saclay, CNRS, Institut d’Astrophysique Spatiale, 91405, Orsay, France 
             }
   \date{Received 24 May 2024; accepted 13 December 2024}

 
  \abstract
   {Radial drift of icy pebbles can have a large impact on the chemistry of the inner regions of protoplanetary disks, where most terrestrial planets are thought to form. Disks with compact mm dust emission ($\lesssim$50 au) are suggested to have a higher \ce{H2O} flux than more extended disks, as well as show excess cold \ce{H2O} emission, likely due to efficient radial drift bringing \ce{H2O}-rich material to the inner disk, where it can be observed with IR facilities such as the \textit{James Webb} Space Telescope (JWST).}
   {We present JWST MIRI/MRS observations of the disk around the low-mass T Tauri star CX Tau (M2.5, 0.37 M$_\odot$) taken as a part of the Mid-INfrared Disk Survey (MINDS) GTO program, a prime example of a drift-dominated disk based on ALMA data. In the context of compact disks, this disk seems peculiar: the source possesses a bright \ce{CO2} feature instead of the bright \ce{H2O} that could perhaps be expected based on the efficient radial drift. We aim to provide an explanation for this finding in the context of radial drift of ices and the disk's physical structure.}
   {We model the molecular features in the spectrum using local thermodynamic equilibrium (LTE) 0D slab models, which allows us to obtain estimates of the temperature, column density and emitting area of the emission. }
   {We detect molecular emission from \ce{H2O, ^12CO2, ^13CO2, C2H2, HCN}, and OH in this disk, and even demonstrate a potential detection of \ce{CO^18O} emission. Analysis of the \ce{^12CO2} and \ce{^13CO2} emission shows the former to be optically thick and tracing a temperature of $\sim${450} K at an (equivalent) emitting radius of $\sim$0.05 au. The optically thinner isotopologue traces significantly colder temperatures ($\sim${200} K) and a larger emitting area. Both the ro-vibrational bands {of \ce{H2O}} at shorter wavelengths and {its} pure rotational bands at longer wavelengths are securely detected. Both sets of lines are optically thick, tracing {a similar temperature of $\sim$500-600 K and emitting area as the \ce{CO2} emission}. We also find evidence for an even colder, $\sim${200} K \ce{H2O} component at longer wavelengths, which is in line with this disk having strong radial drift. We also {find evidence of} highly excited rotational OH {emission} at 9-11 $\mu$m, known as `prompt emission', caused by \ce{H2O} photodissociation. Additionally, we firmly detect 4 pure rotational lines of \ce{H2}, {which show evidence of extended emission}. Finally, we also detect several H recombination lines and the [Ne II] line. }
   {The cold temperatures found for both the \ce{^13CO2} and \ce{H2O} emission at longer wavelengths indicate that radial drift of ices likely plays an important role in setting the chemistry of {the inner disk of CX Tau}. Potentially, the \ce{H2O}-rich gas has already advected onto the central star, which is now followed by an enhancement of comparatively \ce{CO2}-rich gas reaching the inner disk, explaining the enhancement of \ce{CO2} emission in CX Tau. {The comparatively weaker \ce{H2O} emission can be explained by the source's low accretion luminosity. Alternatively, the presence of a small, inner cavity with a size of roughly 2 au in radius, outside the \ce{H2O} iceline, could explain the bright \ce{CO2} emission.} Higher angular resolution ALMA observations are needed to test this.}


   \maketitle
%
\nolinenumbers
\section{Introduction} \label{sec:intro}
The chemistry of the inner few au of protoplanetary disks is an important factor in setting the atmospheric composition of exoplanets. Most planets are thought to form in this region and thus their bulk elemental inventory is set by that of their parent disk \citep{dawson2018, oberg2021}. The hot, inner regions of the disk are best characterized in the infrared (IR). This can now be done in greater detail than ever before with the \textit{James Webb} Space Telescope (JWST) \citep[e.g.][]{kospal2023, banzatti2023, kamp2023, vandishoeck2023, henning2024, pontoppidan2024, romero-mirza2024_sample}, its higher sensitivity and higher spectral resolution even enabling the first detection of \ce{^13CO2} in a T Tauri disk \citep{grant2023}, as well as a large variety of hydrocarbons in disks around very low-mass stars \citep{tabone2023, arabhavi2024, kanwar2024_sz28}. This shows the capability of JWST to uncover the chemistry of these inner regions in unprecedented detail.\\
\newline
The chemical composition of the disk is linked closely to the disk's physical structure, especially to the transport (or lack thereof) of dust grains. In the outer disk, these dust grains contain an icy mantle composed of several frozen-out species, such as \ce{H2O, CO2, NH3}, and CO. When these dust grains drift inwards and cross each species' iceline, the sublimating ice can enrich the gas, changing the elemental composition of the disk with radius, which can be traced by, e.g., the C/O ratio \citep{oberg2011, booth2017}. The C/O ratio in the disk changes not only with radius, but also with time. This is demonstrated by \citet{kalyaan2021, kalyaan2023} and \citet{mah2023,mah2024}, who show that the delivery of O-rich pebbles from just outside the \ce{H2O} iceline will first lead to a decrease of the C/O ratio in the innermost part of the disk. The subsequent draining of this O-rich gas onto the star, combined with the delivery of more C-rich gas advecting inwards from the outer disk, will increase this ratio again over time. \citet{krijt2018, krijt2020} show a similar effect in their modeling work, where the sequestration of species such as CO in ices can lead to a depletion in the gas-phase abundance higher-up in the disk, thus changing the C/O ratio. \citet{kalyaan2021, kalyaan2023} also demonstrate how gaps at different locations in the disks can affect this process, similar to the works by \citet{bitsch2023} and \citet{lienert2024} who investigate the effects of gap-opening on the C/O ratio caused by a giant planet and internal photo-evaporation, respectively. Work by \citet{sellek2024} also demonstrates how disks can undergo a \ce{CO2}-rich phase with time, which is sensitive to the location of a possible gap and seems to be traced well by the \ce{CO2} column density.  \\
\newline
The Atacama Large Millimeter/sub-millimeter Array (ALMA) shows that many disks have substructures in the dust out to large scales \citep[see, e.g.,][]{andrews2018, huang2018, long2018}, which could halt this drift and subsequent enrichment of the gas \citep{kalyaan2021, kalyaan2023}. However, this ubiquity of substructures seems to be limited to the brightest and largest disks, as larger surveys seem to instead find most disks to be very compact, with no discernible substructures \citep{ansdell2018, cieza2019, long2019}. This could be a result of the initial collapsing cloud core having a low angular momentum, allowing it to form a small, compact disk. However, dust grains are also known to drift radially inwards very rapidly, possibly making the dust disk more compact over time. \\
\newline
To distinguish between these two scenarios, a good method is to compare the mm dust disk size to the gas disk size as traced by CO emission. After all, the gas is not subject to the same efficient radial drift as the dust, so it is expected to retain its initial size, which would be much larger than the mm dust disk. In the case of low initial angular momentum, the gas and dust disks would be expected to have a more similar size. Modeling work shows that radial drift can indeed be the cause of (sufficiently large; {$R_{\rm gas}/R_{\rm dust}$}$\gtrsim$4) differences {between} the dust disk extent and the gas disk extent \citep[see][]{facchini2017, trapman2019, trapman2020}.\\
\newline
One of the most extreme examples of a drift-dominated disk found to date is CX Tau, a disk around an M2.5 star in the Taurus-Auriga star-forming region at a distance of 127.9 $\pm$ 0.7 pc \citep{gaia2018}. High-resolution ALMA observations ($\sim$40 mas, corresponding to $\sim$5 au at 128 pc) find a 68\% dust disk radius of only $\sim$15 au, whereas the 68\% gas disk radius in \ce{^12CO} extends out to $\sim$75 au, yielding $R_{\rm gas}/R_{\rm dust} \sim 5$ \citep{facchini2019}, a difference that can only be achieved through very efficient radial drift.\\
\newline
Naturally, one would expect that the chemistry of the inner regions differs between a compact disk with efficient radial drift and a large disk with substructures halting this process. This is suggested to be the case by \citet{banzatti2020}, who find a correlation between the \ce{H2O} flux observed with \textit{Spitzer} and the dust disk size, with the smaller disks being more \ce{H2O}-rich. This correlation is also theoretically motivated with modeling work by \citet{kalyaan2021}. The larger \ce{H2O} flux in compact disks is attributed to the efficient radial drift in compact disks delivering \ce{H2O}-rich ices to the inner regions. More recent work by \citet{banzatti2023} also shows that compact disks show excess cool \ce{H2O} emission, which can likely be attributed to efficient radial drift as well.\\
\newline
We present observations of CX Tau taken with the Mid-InfraRed Instrument \citep[MIRI;][]{rieke2015, wright2015, wright2023} Medium Resolution Spectrometer \citep[MRS;][]{wells2015, argyriou2023} on board JWST, taken as part of the MIRI Mid-INfrared Disk Survey (MINDS) GTO program (PID: 1282, PI: T. Henning; see \citealt{henning2024} for an overview of the program). The central star has a luminosity of $L_* = 0.22 \pm 0.11\; L_\odot$, an effective temperature $T_{\rm eff} = 3483 \pm 48$ K \citep{herczeg2014}, a dynamical estimate of its mass of $M_* = 0.37\; M_\odot$ \citep{simon2017}, and a mass accretion rate of $\dot{M}_{\rm acc} = 7.1\times10^{-10}\; M_\odot\;{\rm yr}^{-1}$ \citep{hartmann1998}. The source was also observed with \textit{Spitzer} \citep{furlan2006, najita2007, furlan2011}, but its molecular features have not been analyzed before. We detect mid-IR emission  from \ce{H2O, ^12CO2, ^13CO2, C2H2, HCN}, and OH for the first time in this disk, and even present a potential detection of \ce{CO^18O}. We use local thermodynamic equilibrium (LTE) slab models to constrain the properties of the emission and use the relative strengths of the emission features to test various scenarios. \\
\newline
This work is structured as follows: in Sect. \ref{sec:methods} we discuss the reduction of our data and our slab-modeling procedure. In Sect. \ref{sec:results} we present the main results of our slab model fits and other analysis, and in Sect. \ref{sec:discussion} we discuss several possible scenarios that could explain our finding of bright \ce{CO2} and faint \ce{H2O} emission. We present our main conclusions in Sect. \ref{sec:conclusions}.

\section{Methods}\label{sec:methods}

\subsection{Observations and data reduction}\label{subsec:obs}
CX Tau was observed with MIRI/MRS on February 20, 2023. The exposures were taken using a four-point dither pattern in the positive direction with a total on-source integration time of 1 hour. The data cover a wavelength range from 4.9 to 27.9 $\mu$m, over which the spectral resolving power $R$ ranges from approximately 3500 to 1500 \citep{argyriou2023, jones2023}. The observations were reduced using the MINDS pipeline \citep{MINDSpipeline}, which combines the standard JWST pipeline \citep[version 1.13.4,][]{bushouse2023} and routines from the VIP package \citep{gomezgonzales2017, christiaens2023}. The pipeline is structured around three main stages that are the same as in the JWST pipeline, namely Detector1, Spec2 and Spec3. \\
\newline
We made use of the bad pixel correction routines contained in the VIP package. This routine identifies the bad pixels through a sigma-filtering method and corrects them using a Gaussian kernel. Centroiding of the central source was also carried out using routines from the VIP package. In particular, {we aligned} the spectral cubes through an image cross-correlation routine, followed by a 2D Gaussian fit in the weighed mean image of each cube. These centroids are used for the aperture photometry, where we extracted a spectrum by summing the signal in a 2.0 full width at half maximum (FWHM = 1.22 $\lambda/D$ with $D$=6.5 m) aperture centered on the source. To ensure that background emission is not affecting the spectrum, the background was estimated from an annulus surrounding the used aperture. We accounted for the flux of the PSF outside the aperture and inside the annulus using aperture correction factors \citep{argyriou2023}. Finally, we corrected the spectrum for residual fringing using the residual fringe correction step in the Spec3 stage. The observations were taken without target acquisition, which may have caused the final spectrum to be slightly more affected by residual fringing even after this final step than it would otherwise be.\\
\begin{figure*}[t]
    \centering
    \makebox[\textwidth][c]{\includegraphics[width=\textwidth]{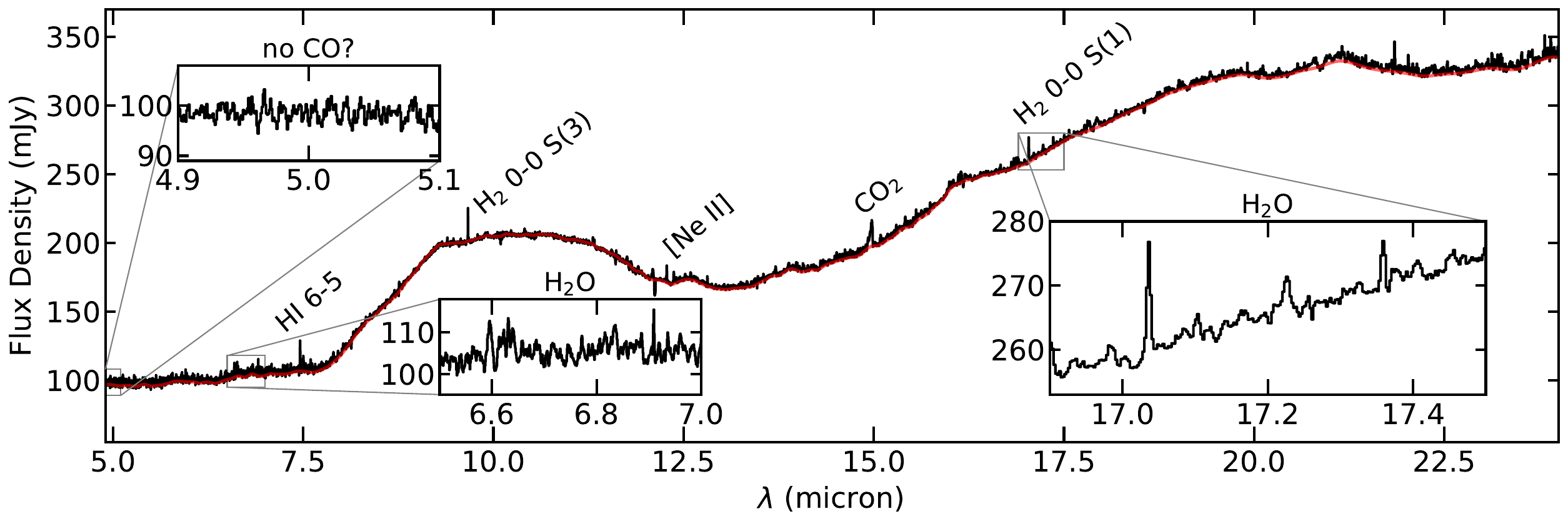}}
    \caption{The full JWST-MIRI MRS spectrum of CX Tau. Several emission features are labeled or shown in insets, and the continuum fit is shown in red.}
    \label{fig:full_spect}
\end{figure*}
\newline
We subtracted our continuum using the method described in \citet{temmink2024a}. This method first identifies and masks downwards spikes using an outlier detection technique involving a Savitzky-Golay filter. The continuum is subsequently subtracted using the 'Iterative Reweighted Spline Quantile Regression' (IRSQR) implementation of the python-package \textsc{pybaselines} \citep{pybaselines}. We have placed the spline knots every 25 data points and used a quantile value of {0.1}.

\subsection{Slab modeling procedure}\label{subsec:slab_models}

The molecular emission present in the spectrum is fit with 0D LTE slab models. The procedure used in this work is the same as that described in \citet{kamp2023}. We refer the reader to that work for an in-depth explanation, and provide only a summary of the most important details here. \\
\newline
The LTE slab models aim to reproduce the data using 3 free parameters: the line-of-sight column density $N$, the gas temperature $T$, and an emitting area $A$. {This emitting area can be converted to an equivalent slab radius assuming $A = \pi R^2$. This slab radius can be equal to the true emitting radius of the emission, but can also reflect the emission originating from an annulus further out in the disk instead.} The line profiles are assumed to be Gaussian, and following \citet{salyk2011} have a full width at half maximum of $\Delta V=4.7$ km s$^{-1}$. \\
\newline
We used slab models to fit emission from \ce{^12CO2, H2O, C2H2, ^13CO2, OH}, and HCN, in that order. The emission from each species was fit to the data individually, after which it was subtracted to fit the emission from the next species. This process is demonstrated in the Appendix in Figs. \ref{fig:slab_15um_ind_p1} and \ref{fig:slab_15um_ind_p2}. 
Since \ce{CO2, C2H2} and HCN all have $Q$ branch features (very densely packed lines) in the MIRI wavelength range, we accounted for mutual shielding of adjacent lines for these species individually, as is described in \citet{tabone2023}. \\
\begin{figure*}
    \centering
    \includegraphics[width=\textwidth]{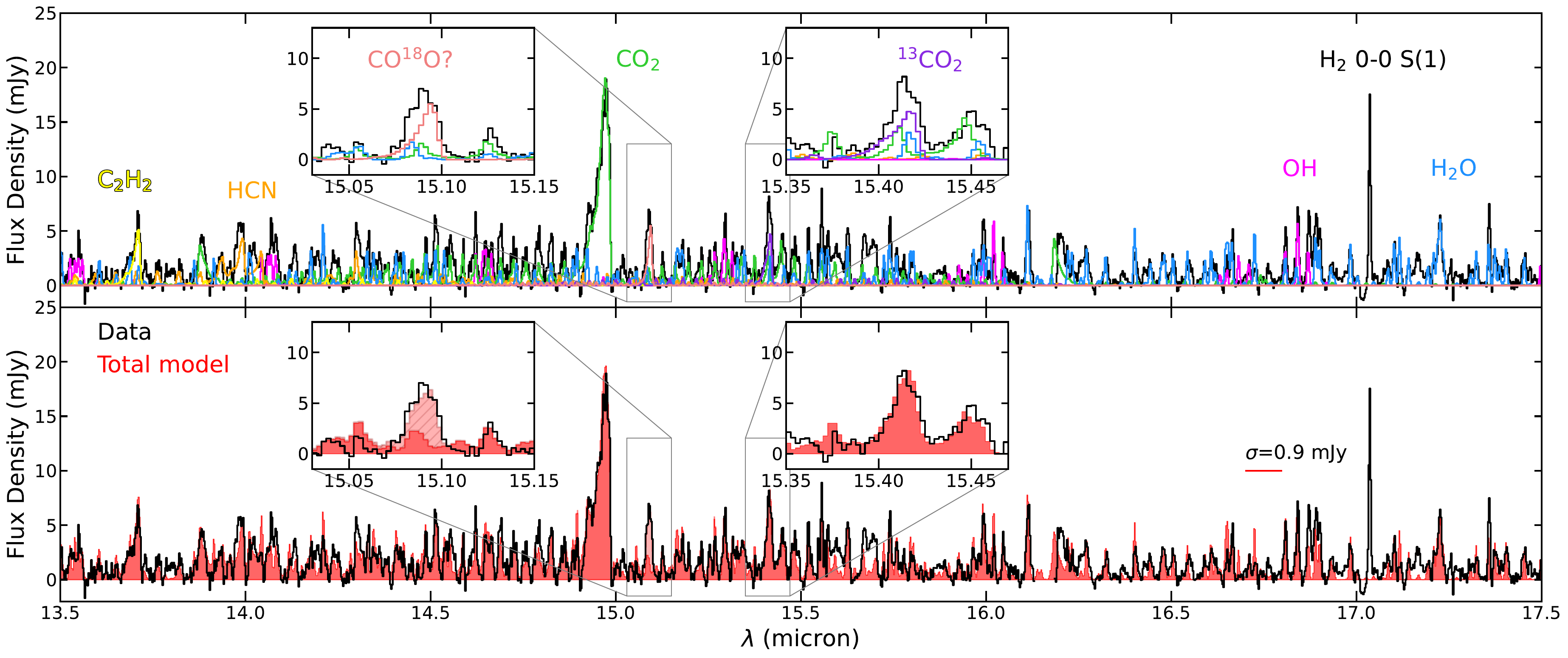}
    \caption{A zoom-in of the 13.5-17.5 $\mu$m region of the CX Tau spectrum (black), together with all slab model fits to the molecular features. These models include \ce{C2H2} (yellow), HCN (orange), \ce{^12CO2} (green), \ce{^13CO2} (purple), \ce{H2O} (blue) and OH ({magenta}). {In light pink, we show a slab model demonstrating a potential detection of \ce{CO^18O}.} The bottom panel shows the data in black, the combined model with and without the \ce{CO^18O} model in light pink and red, respectively. The red bar indicates the window in which the uncertainty was estimated. {An artifact around 16.15 $\mu$m has been masked.} The insets show close-ups of the regions around the \ce{^13CO2} emission feature and the (potential) \ce{CO^18O} feature.} 
    \label{fig:slab_15um}
\end{figure*} 
\newline
We fit these models in the 13.5-17.5 $\mu$m region, where most molecular features are located. For \ce{H2O}, we perform two separate fits in two different wavelength regions (5.5-8.5 $\mu$m and 13.5-17.5 $\mu$m), as these wavelength ranges are expected to probe different regions in the disk, and thus their excitation conditions are expected to differ \citep[see, e.g.,][]{gasman2023b, banzatti2023}. All slab models were convolved to the average resolving power within the fitted region (corresponding to $\lambda/\Delta\lambda \approx 3500$ at the shortest wavelengths and to $\lambda/\Delta\lambda \approx 1500$ at the longest wavelengths), after which they were resampled to the wavelength grid of that region using the SpectRes package \citep{carnall2017}. The best-fit model was obtained from a grid of models using a $\chi^2$ method as described in \citet{grant2023}, \citet{gasman2023b}, and \citet{schwarz2024}. The spacing of this grid was kept consistent for most fits, where $T$ was varied linearly between 100 and 1000 K in steps of $\Delta T=22.5$ K, $N$ was varied in log-space between $10^{14}$ and $10^{23}$ cm$^{-2}$ with steps of $\Delta \log N = 0.225$. {Exceptions were made for \ce{^13CO2}, for which $T$ was varied between 50 to 500 K in steps of $\Delta T = 11.25$ K, and for OH, for which $T$ was varied between} 1000 and 3000 K in steps of $\Delta T = 50$ K. For HCN, we restricted $N$ between $10^{14}$ and $10^{21}$ cm$^{-2}$ as the fit is very unconstrained and gave unrealistically high values otherwise. For each point in the grid, the best-fit $R$ is determined by minimizing the {reduced} $\chi^2$. The $\chi^2$ fit is performed in specific spectral windows that contain important, characteristic features of the emission, avoiding contamination from other species. The average rms $\sigma$ was calculated from the spectrum itself by taking the standard deviation of a relatively line-free region, for which the best-fit slab models were first subtracted. The fits to the OH and HCN emission were found to be unconstrained, and the uncertainties on the other fits are generally large. This is further discussed in Sect. \ref{subsec:molecular} and Appendix \ref{app:slab}.

\section{Results} \label{sec:results}
\subsection{Full spectrum} \label{subsec:full_spect}

The full spectrum of CX Tau is presented in Fig. \ref{fig:full_spect}. The shape of the spectrum is as expected for a typical T Tauri disk: a gradual increase in flux towards the longer wavelengths, with a strong silicate feature around 10 $\mu$m. The shape of this feature has been shown to be sensitive to the average grain size in the disk, where the feature becomes broader and more flat-topped as the grains in the disk grow larger \citep[see, e.g.][]{bouwman2001, przygodda2003, kessler-silacci2006}. As such, the silicate feature in this spectrum indicates that the dust in CX Tau is more evolved, having likely already grown to larger sizes. Additionally, several smaller ``bumps'' can be seen in the spectrum around 12.5, 16, 19, and 21 $\mu$m. These features are most likely caused by emission from crystalline forsterite \citep{olofsson2009, olofsson2010}.

\subsection{Molecular emission} \label{subsec:molecular}

The most prominent molecular feature in the spectrum of CX Tau {presented in Fig. \ref{fig:full_spect}} is the \ce{CO2} $Q$ branch at 15 $\mu$m. All other molecular features clearly have lower line-to-continuum ratios and as such they were likely unable to be detected with \textit{Spitzer}, requiring the increased sensitivity and spectral resolution of MIRI. We detect weak \ce{H2O} emission across the full spectrum, though the ro-vibrational bands at the shorter wavelengths ($\sim$5-9 $\mu$m) are slightly more prominent than the pure-rotational lines at the longer wavelengths ($\sim$14-24 $\mu$m). Features of \ce{^13CO2, C2H2}, and HCN are also detected, and close inspection of the 13.5-17.5 $\mu$m region also reveals a potential detection of \ce{CO^18O} (Fig. \ref{fig:slab_15um} and below). {OH emission is clearly detected at longer wavelengths ($\sim$14-24 $\mu$m) where the lines are produced by chemical pumping through the \ce{O + H2 -> OH + H} reaction and collisional excitation. We also find evidence of extremely excited rotational lines at shorter wavelengths (9-11 $\mu$m), known as prompt emission produced by \ce{H2O} photodissociation.} {Emission from the high-$J$ ro-vibrational transitions of \ce{CO} at $\sim$5 $\mu$m is not detected above 3$\sigma$. This is in line with the non-detection of CO emission reported by \citet{anderson2024} using Keck/NIRSPEC, who instead detect narrow blue-shifted absorption indicative of a photo-evaporative wind.} Finally, we also detect prominent emission from the pure-rotational transitions of \ce{H2}{, which show extended emission,} as well as atomic emission from several hydrogen recombination lines and the forbidden [Ne II] line {(see Appendix \ref{app:ext})}. \\
\newline
To analyze the molecular emission, LTE slab model fits are used to provide an indication of the emission's temperature, column density and emitting area (Sect. \ref{subsec:slab_models}). We fit these models for the features from \ce{^12CO2}, \ce{^13CO2, H2O, C2H2, HCN}, and OH. {Due to the weakness of most molecular features and the relatively low S/N of this spectrum, most of these emission properties are quite uncertain. For comparison, the \ce{H2O} lines in CX Tau have fluxes that are $\sim$100-150 times weaker than those detected in DR Tau \citep{temmink2024b, temmink2024a}. As such, we focus our analysis on detections rather than constraining parameters, and we only report on emission properties when they can be deemed robust. We discuss this further in Sect. \ref{app:fit}.} The individual fits are shown in Appendix \ref{app:slab} in Figs. \ref{fig:slab_15um_ind_p1}, \ref{fig:slab_15um_ind_p2} and \ref{fig:slab_15um_ind_closeups} and the $\chi^2$ maps {of the constrained fits} are  shown in Figs. \ref{fig:chi2_15um}. 
\begin{figure}
    \centering
    \includegraphics[width=0.5\textwidth]{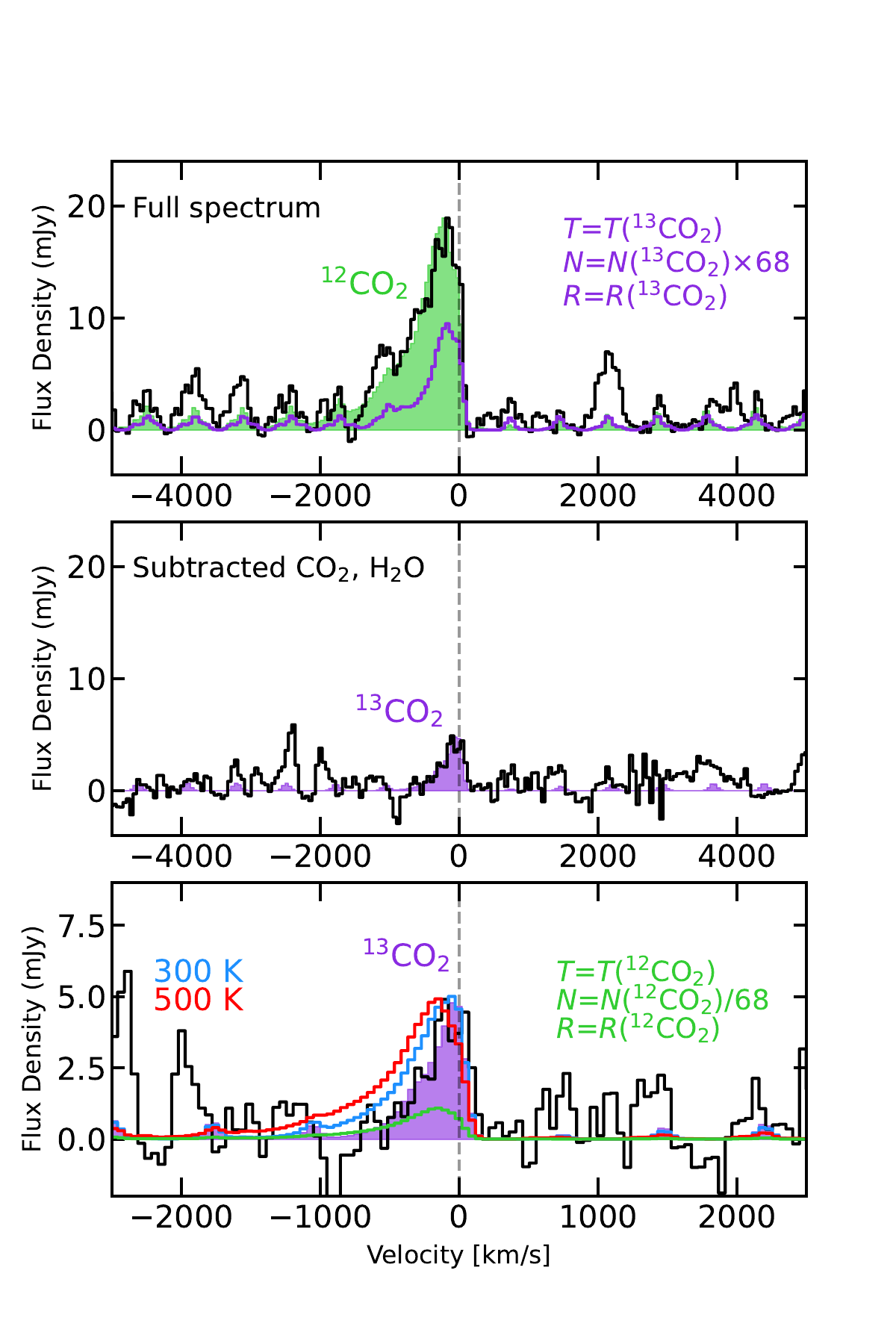}
    \caption{A comparison of the \ce{^12CO2} (green) and \ce{^13CO2} (purple) $Q$ branch shapes. Top: {a zoom-in of the \ce{^12CO2} $Q$ branch with the best-fit slab model plotted in the green shaded region. A model {with} the derived \ce{^13CO2} parameters is plotted in purple (see text).} Middle: a zoom-in of the \ce{^13CO2} $Q$ branch on the same vertical scale as {the top panel}, where the emission from \ce{^12CO2} and \ce{H2O} has been subtracted. The best-fit slab model is plotted in the purple shaded region. Bottom: a further zoom-in of the middle panel {where the best-fit model (180 K;} purple shaded region) {and slab models of} 300 K (blue line) and 500 K (red line) are shown. The latter two models have their emitting radius scaled to {produce the same peak flux}. In green, a slab model {with} the derived \ce{^12CO2} parameters (see text) is shown.}
    \label{fig:slab_13CO2}
\end{figure}

\subsubsection{\ce{^12CO2}, \ce{^13CO2} and \ce{CO^18O}}\label{subsub:CO2}
The $Q$ branch of \ce{CO2} at 15 $\mu$m is the most prominent molecular feature detected in this source (Figs. \ref{fig:full_spect} and \ref{fig:slab_15um}).{The fit to the \ce{CO2} emission indicates} that the emission is tracing warm ($\sim${450} K), optically thick gas with a column density of $\sim$$8\times10^{17}$ cm$^{-2}$ {at an (equivalent) emitting radius of $\sim$0.05 au}. The optically thinner isotopologue, \ce{^13CO2}, is also firmly detected. The fit to this feature is less well-constrained than that for \ce{^12CO2}, but it is clear from the $\chi^2$ plot (see Fig. \ref{fig:chi2_15um}) that the emission from \ce{^13CO2} could still be marginally optically thick with its best-fit column density of $\sim$$2\times10^{17}$ cm$^{-2}$ (though this value is not very well-constrained). We find the emission to be very cold ($\sim${200} K), and it traces a larger emitting area {of $\sim$0.2 au}. {If we assume that the \ce{^13CO2} emission is a more reliable tracer of the total \ce{CO2} column density, we find $N$(\ce{CO2}) = $68\times N$(\ce{^13CO2}) $\approx 10^{19}$ cm$^{-2}$.} \\
\newline
When investigating the residuals of the fit of the \ce{^12CO2, ^13CO2, H2O, C2H2, HCN}, and OH emission to the 13.5-17.5 $\mu$m region, we find a feature larger than 3$\sigma$ at 15.07 $\mu$m (see the bottom panel of Fig. \ref{fig:slab_15um_ind_p1} and Sect. \ref{app:residuals}). This is particularly interesting as it coincides with the $Q$ branch of \ce{CO^18O}. Given the bright \ce{^12CO2} emission detected in CX Tau, combined with the detection of \ce{^13CO2}, a contribution from this rarer isotopologue to the observed emission could certainly be expected. We demonstrate in Fig. \ref{fig:residuals_15} that there is evidence of \ce{CO^18O} contributing to the feature seen at 15.07 $\mu$m, as it seems too strong to be attributed solely to \ce{H2O} emission. We discuss this further in Sect. \ref{app:residuals}. We only consider this a potential detection and we thus do not fit the emission with a $\chi^2$ fit as we do for the other species in this wavelength range. \\
\newline
In Fig. \ref{fig:slab_15um}, we show a \ce{CO^18O} model that provides a good match to the data. We assume a column density of $5\times10^{16}$ cm$^{-2}$, assuming $N$(\ce{^13CO2})/$N$(\ce{CO^18O}) $\sim$ {4} (from the ISM abundance ratios of \ce{^12C}/\ce{^13C} $\sim 68$ and \ce{^16O}/\ce{^18O} $\sim 500${, accounting for the fact that \ce{CO2} has 2 O atoms}; \citealt{wilson1994, wilson1999, milam2005}). The feature is quite narrow, implying a low temperature (see Fig. \ref{fig:residuals_15}), which is in line with our finding of a low temperature for \ce{^13CO2}. The insets in Fig. \ref{fig:slab_15um} demonstrate the improvement that is made to the total fit when the contribution from \ce{CO^18O} is included. We also see that the \ce{^13CO2} and \ce{CO^18O} features are similar in strength. This is in line with findings of the eXtreme UV Environments (XUE) JWST GO program (PI: Ram\'irez-Tannus; \citealt{ramirez-tannus2023}) , who detect these species (as well as \ce{CO^17O}) in an externally irradiated disk (Frediani et al. in prep.). The \ce{CO^17O} isotopologue is found to be the weakest of the three detected isotopologues in that work, so our data likely lack the S/N to detect it.\\
\newline
We investigate the difference in temperature between the \ce{^12CO2} and \ce{^13CO2} emission further in Fig. \ref{fig:slab_13CO2}. {This difference} is especially evident when looking at the shape of their respective $Q$ branches. The data are shown as a function of velocity, centered on the $Q$(8) line. This figure clearly demonstrates the sensitivity of these features to the temperature \citep[see also][]{bosman2017, grant2023}, as the \ce{^12CO2} feature is clearly broader and its peak is blueshifted compared to the \ce{^13CO2} feature ({which is shown in the middle and bottom panels, where the emission from \ce{^12CO2} and \ce{H2O} is subtracted to isolate the feature}). The bottom panel of Fig. \ref{fig:slab_13CO2} {compares the} \ce{^13CO2} $Q$ branch with several slab models. In purple, the best-fitting slab model with $T$ = {180} K is shown, and models with the same column density but $T$ = 300 and 500 K (scaled with $R$ to have the same peak flux) are shown in blue and red, respectively. This shows that the temperature of the \ce{^13CO2} emission can really be constrained to {<300} K, as the 300 K model is already too broad. \\
\newline
Finally, {it is interesting to} test whether the \ce{^12CO2} and \ce{^13CO2} are tracing the same gas. {If this were the case, we should be able to reproduce the \ce{^13CO2} feature with a slab model that has a temperature and emitting radius corresponding to the best-fit parameters of our \ce{^12CO2} model, and a column density that is 68 times smaller, following the ISM \ce{^12C /^13C} ratio. The model in question is shown in green in the bottom panel of Fig. \ref{fig:slab_13CO2}.} Clearly, this model does not reproduce the observed \ce{^13CO2} feature. {This thus indicates that the observed \ce{^12CO2} and \ce{^13CO2} emission are not tracing the same gas, likely due to the \ce{^12CO2} emission being optically thick. We also show the same test in reverse in the top panel of Fig. \ref{fig:slab_13CO2}, where we show a \ce{^12CO2} model with the temperature and emitting radius of our best-fitting \ce{^13CO2} model, and with a column density 68 times larger. This model is shown in purple, and also does not reproduce all of the observed \ce{^12CO2} emission.}\\
\begin{figure*}
    \centering
    \makebox[\textwidth][c]{\includegraphics[width=1\textwidth]{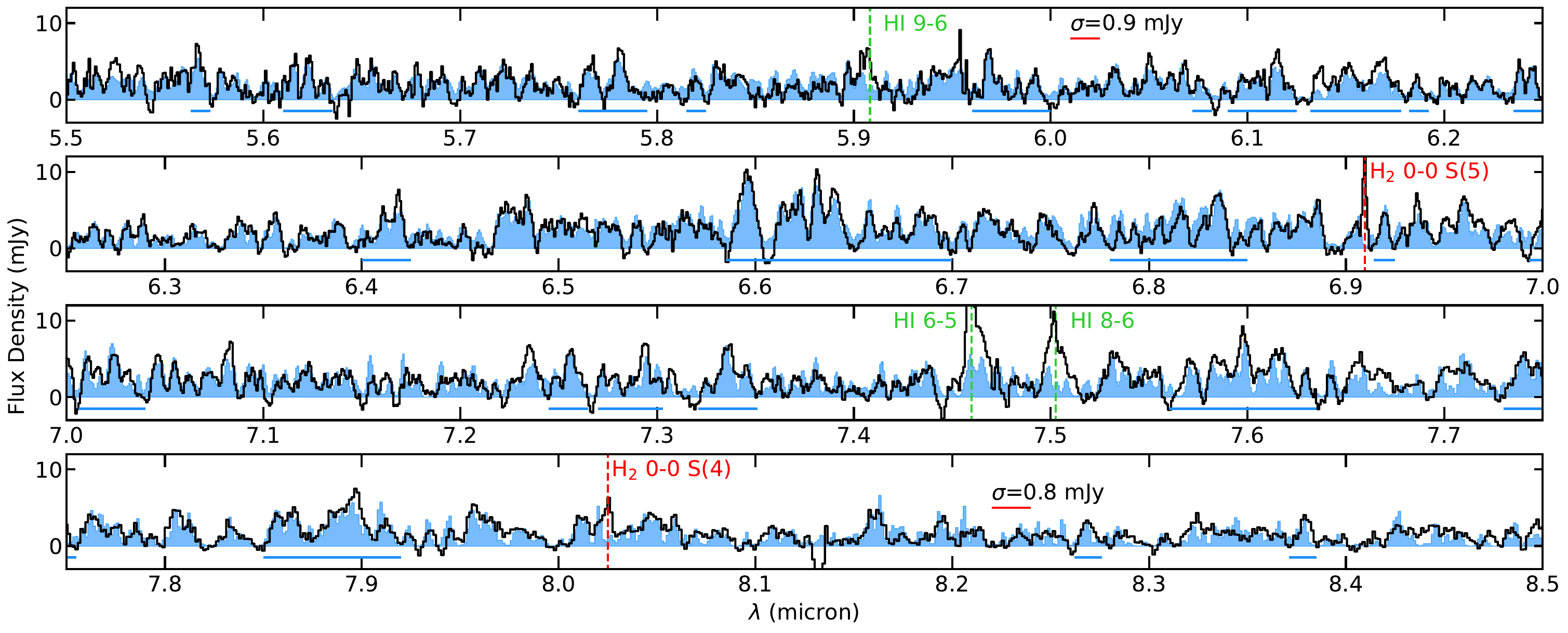}}
    \caption{{Four} panels showing a zoom-in of the 5.5-8.5 $\mu$m region of the CX Tau spectrum (black), together with the \ce{H2O} slab model fits (blue). The {four} different panels show the regions which were used to perform the \ce{H2O} $\chi^2$ fits with horizontal blue bars.  }
    \label{fig:slab_H2O_short}
\end{figure*}
\newline
{Thus, Fig. \ref{fig:slab_13CO2} demonstrates two things: the \ce{^13CO2} emission is colder than the \ce{^12CO2} emission and the two do not trace the same gas.} This finding brings forth two possible explanations. The most straightforward explanation is that the \ce{^13CO2} emission is more optically thin, {tracing} deeper layers {of the disk} and thus lower temperatures. As the CX Tau disk has a moderate inclination (55\degree), this also means the emission traces further out into the disk radially, potentially explaining why the emitting radius for \ce{^13CO2} is found to be larger than that for \ce{^12CO2}. In this context, the potential detection of \ce{CO^18O} could also be extremely valuable, as this rarer isotopologue is even more optically thin. This could thus allow us to constrain the total \ce{CO2} column density even better, especially as the $\chi^2$ fit of the \ce{^13CO2} emission retrieves a relatively high column density and indicates that its emission may still be marginally optically thick.\\
\newline
However, the cold temperature of the \ce{^13CO2} emission could also indicate that it traces emission close to the \ce{CO2} snowline, possibly enhanced by radial drift and sublimating ices \citep[see, e.g.][]{bosman2017}. This cold component near the snowline would then also provide a part of the \ce{^12CO2} emission that is observed. This {may be quantified by the purple slab model shown in the top panel of Fig. \ref{fig:slab_13CO2}, which only reproduces part of the observed \ce{^12CO2} emission. Thus, the emission may be made up of both a colder and hotter component, much like what is found for \ce{H2O} \citep[e.g.][]{banzatti2023, temmink2024b, romero-mirza2024_as209}. Our data lack the S/N to investigate this much further, but it} motivates the study of \ce{^12CO2}, \ce{^13CO2} and \ce{CO^18O} emission with a more complex temperature structure {in future work}.

\subsubsection{\ce{C2H2} and HCN}\label{subsub:HCN}

As can be seen in Fig. \ref{fig:slab_15um}, emission from \ce{C2H2} and HCN is also detected in the 13.5-17.5 $\mu$m region. Since the emission from these molecules is {much} weaker than that of \ce{CO2}, the {emission properties retrieved from our} fits are {quite uncertain}. The emission from \ce{C2H2} is blended with \ce{H2O} emission, but still detected. The fit indicates that the emission {is warm and may be} optically thin, {and thus the column density and emitting area are poorly constrained due to their degeneracy}. The HCN fit is the least constrained due to significant blending with \ce{H2O}, \ce{CO2}, and OH. {While the emission is likely detected, its emission properties cannot be reliably determined by our fit}.

\begin{figure*}
    \centering
    \makebox[\textwidth][c]{\includegraphics[width=1\textwidth]{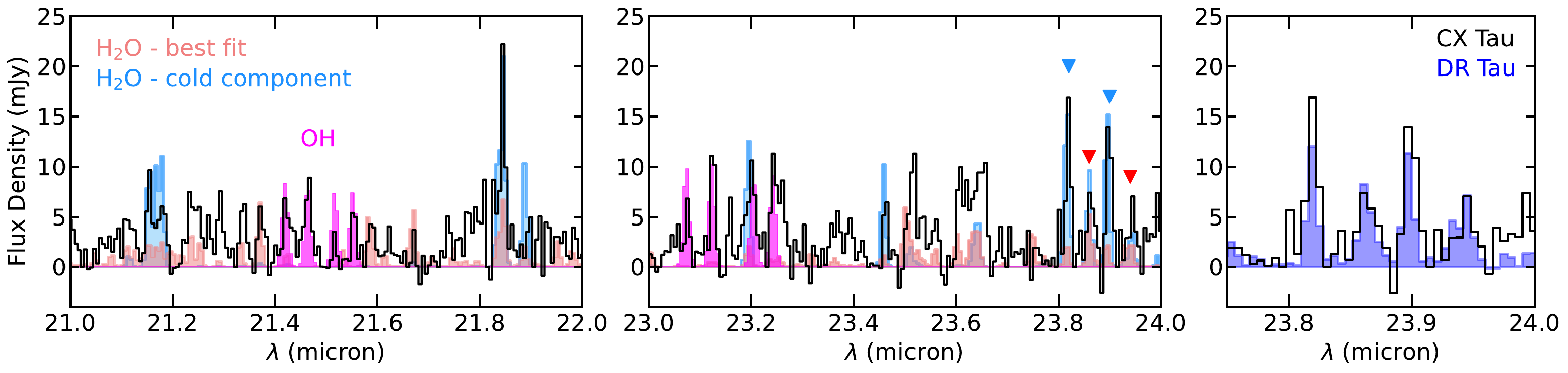}}
    \caption{Three panels showing a zoom-in of the 21-22 $\mu$m region, the 23-24 $\mu$m region, and the 23.75-24 $\mu$m region, respectively, of the CX Tau spectrum (black). The left and middle panels show a {warm} \ce{H2O} slab model in light pink {(the best-fit model to the 13.5-17.5 $\mu$m region)}, a {colder (175 K)} slab model in light blue and an OH slab model in magenta. Blue and red triangles denote \ce{H2O} lines that demonstrate the presence of a cold \ce{H2O} component in CX Tau (see text). The right panel shows the spectrum of CX Tau in black and the spectrum of DR Tau \citep{temmink2024b, temmink2024a} in dark blue. The spectrum of DR Tau has been scaled to match the flux of the 23.87 $\mu$m line.}
    \label{fig:slab_H2O_long}
\end{figure*}

\subsubsection{\ce{H2O}}\label{subsub:H2O}

\ce{H2O} emission is weak, but detected. {We fit it }with slab models in {two} wavelength ranges. In the 13.5-17.5 $\mu$m region, we can clearly detect several pure-rotational \ce{H2O} lines around 17.2 $\mu$m, as can be seen in Fig. \ref{fig:slab_15um}. 
The best fit indicates that the \ce{H2O} emission traces gas of a similar temperature ($\sim$500-600 K) and emitting radius ($\sim$0.05 au) as the observed \ce{CO2} emission. Our best fit finds a column density of $\sim$10$^{19}$ cm$^{-2}$, which is also similar to that of \ce{CO2} if one assumes that the \ce{^13CO2} column density provides a better measure of the total \ce{CO2} column. However, if one chooses narrower fit windows (e.g. using the isolated lines suggested by \citealt{banzatti2024}), a lower column density of $\sim$10$^{18}$ cm$^{-2}$ can also be retrieved. There are some weak features in the spectrum that indicate that a higher column density may be preferred, though our best-fit \ce{H2O} column density of $\sim$10$^{19}$ cm$^{-2}$ may also be slightly overestimated as our data lack the S/N to properly constrain this value, and these features could also indicate the presence of hotter \ce{H2O}, rather than a larger column. {This is also implied by the fact that the best fit to the 13.5-17.5 $\mu$m region seems to underproduce the observed emission at shorter wavelengths ($\sim$10-13 $\mu$m; see, e.g., Figs. \ref{fig:H_lines}-\ref{fig:Ne_lines}), indicating the potential need for a hotter component}.\\
\newline
We fit the ro-vibrational emission at the shorter wavelengths separately from the rotational emission at 13.5-17.5 $\mu$m, as these lines are expected to trace different excitation conditions \citep[see, e.g.][]{bosman2022a, banzatti2023_ground}. This fit is presented in Fig. \ref{fig:slab_H2O_short}, showing that almost all of the emission detected at these wavelengths is coming from \ce{H2O}. This model indicates that this emission {has a similar temperature, but is} coming from a {slightly} smaller emitting area compared to the rotational \ce{H2O} emission at 13.5-17.5 $\mu$m. This is similar to what is found in other, more in-depth studies of \ce{H2O} emission \citep[e.g.][]{gasman2023b, banzatti2023_ground, romero-mirza2024_as209, temmink2024b}. {The column density is difficult to constrain due to the low S/N and the retrieved value of $N$ >10$^{20}$ cm$^{-2}$ is likely overestimated by our best fit.} We {also point out} that this emission is likely out of LTE due to the high critical density of these transitions, leading to sub-thermal population of the levels \citep[e.g.][]{meijerink2009, bosman2022a}. Still, our results show that a robust identification of \ce{H2O} can be made. \\
\newline
At longer wavelengths ($>$18 $\mu$m), obtaining a constraint on the emission properties of the \ce{H2O} becomes {even} more difficult, as the S/N ratio decreases {further}. As such, we do not perform a $\chi^2$ fit in this region. We find that the slab model that was fit to the 13.5-17.5 $\mu$m region provides {evidence that warm \ce{H2O} emission is presumably still present in the spectrum} out to $\sim$24 $\mu$m. We also find indications around 21 and 23 $\mu$m for a colder component of $\sim${200} K. This is shown in the left and middle panels of Fig. \ref{fig:slab_H2O_long}, where the former model is shown in light pink and the latter in light blue. \\
\begin{figure*}
    \centering
    \includegraphics[width=\textwidth]{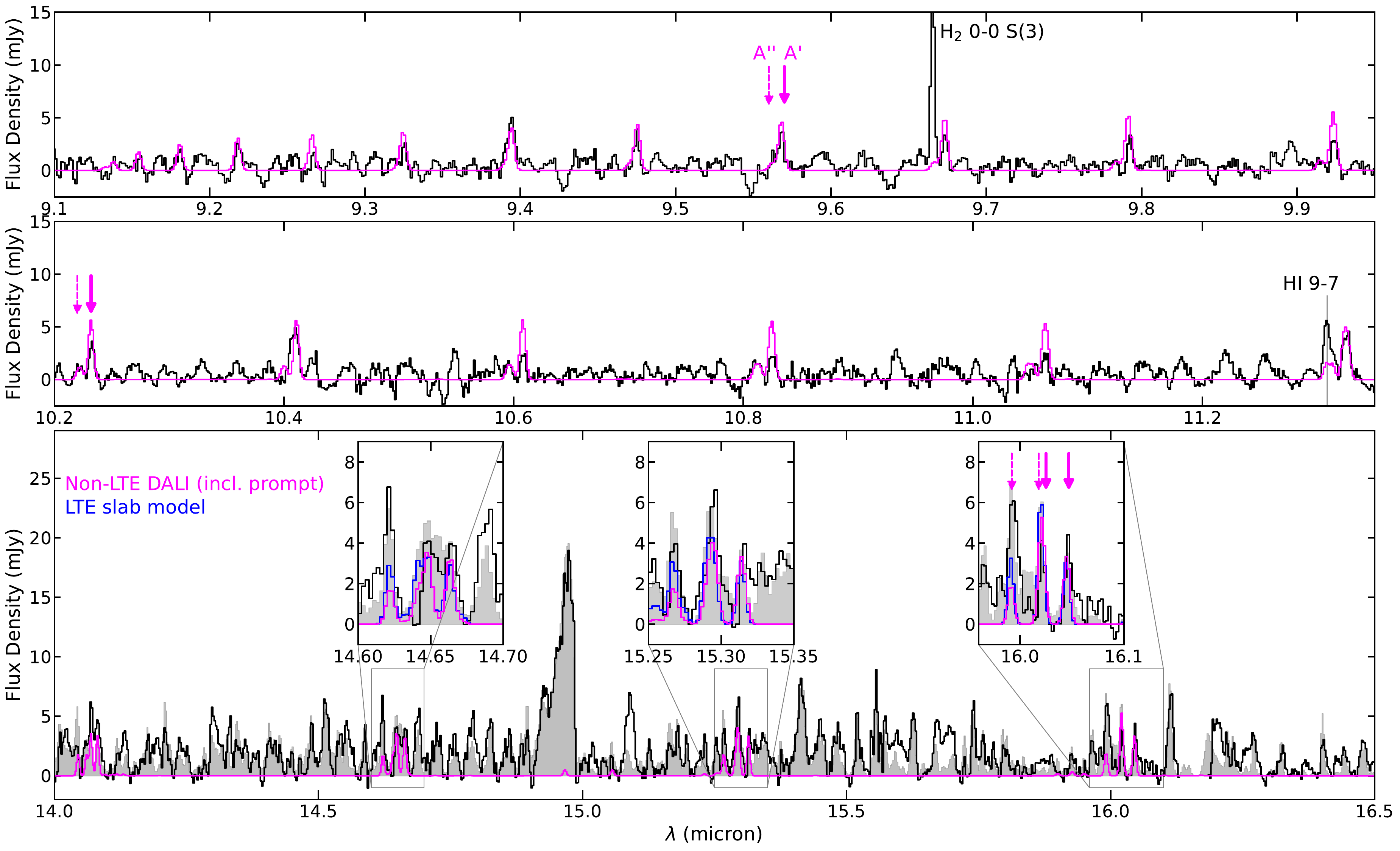}
    \caption{Zoom-ins of the CX Tau spectrum (black) from {9-16.5} $\mu$m, demonstrating the {potential} detection of OH prompt emission. The first two panels show close-ups of the 9-11 $\mu$m region, with a non-LTE DALI model from \citet{tabone2024} ({which includes} both prompt emission due to photodissociation of \ce{H2O} and thermal excitation) shown in pink. The third panel shows the contributions of all slab models fit in the 13.5-17.5 $\mu$m region in grey (including OH). The DALI model is overlayed in pink. {An artifact around 16.15 $\mu$m has been masked.} The insets in this panel show close-ups of this model contrasted with the best-fit OH LTE slab model which is shown in blue. Lines belonging to the A' symmetry are indicated with thick, solid {pink} arrows and lines belonging to the A'' symmetry are indicated with thin, dashed {pink} arrows. }
    \label{fig:OH_prompt}
\end{figure*}
\newline
As demonstrated in \citet{banzatti2023} and \citet{temmink2024b}, the set of lines between 23.8 and 23.9 $\mu$m (indicated with blue and red triangles in the middle panel of Fig. \ref{fig:slab_H2O_long}) provides especially good evidence for the presence of this colder component to the \ce{H2O} emission, as the spectrum shows that the {lines indicated in blue} are brighter than {those indicated in red}.
{The latter have an upper level energy that is twice as large ($\sim$3000 K vs. $\sim$1500 K) and thus they rapidly grow in strength relative to the lower $E_{\rm up}$ lines when the temperature of the slab model is increased.} To further demonstrate this, we show a close-up of this set of lines in the third panel of Fig. \ref{fig:slab_H2O_long} where we compare our data to DR Tau, a disk for which the presence of this cold component has been firmly established \citep{temmink2024b}. From this figure, it is clear that these lines have a similar ratio in CX Tau and thus this cold component is present in its spectrum as well. \\
\newline
The detection of this cold \ce{H2O} component motivates the use of more complex models to better comprehend the data, for example by including multiple temperature components in a slab model fit \citep[e.g., analysis in][]{pontoppidan2024, temmink2024b} or even more complexity in the form of radial column density and temperature gradients \citep[e.g.][]{kaufer2024, romero-mirza2024_sample}.

\subsubsection{OH}\label{subsub:OH}

OH emission is detected at wavelengths beyond 13 $\mu$m, as already shown in Fig. \ref{fig:slab_15um}. The fit in this region is not very well-constrained. The temperature found from the $\chi^2$ fit is very high (>1500 K), which is not representative of the physical gas temperature, but rather an effect of chemical pumping through the \ce{O + H2 -> OH + H} reaction, as is known to affect OH emission in this wavelength range \citep{tabone2021}. We also detect OH emission at longer wavelengths. Due to the lower S/N in this region, we only fit this emission visually to demonstrate {that it is likely also present at these wavelengths}, which can be seen in Fig. \ref{fig:slab_H2O_long}.\\
\newline
Additionally, we {find evidence of} highly excited suprathermal rotational OH {emission} at shorter wavelengths (9-12 $\mu$m). These transitions originate in pure rotational states with rotational quantum numbers $N$ as high as 44, which have upper level energies as high as $\sim$40000 K. These lines are known as `prompt emission' {and originate} from \ce{H2O} photodissociation by photons in the 114-144 nm wavelength range (which includes Ly$\alpha$), which brings the resulting OH molecule into a very high rotational state \citep{vanharrevelt2003, tabone2021}. The resulting cascade down the rotational ladder produces prompt emission. \\
\newline
One unique characteristic of prompt emission by \ce{H2O} photodissociation is that the produced OH will only be excited in two of the four hyperfine states that make up each rotational state \citep{zhou2015}. Each rotational state is split by spin-orbit coupling, labeled by $\Omega = 1/2, 3/2$ and $\Lambda$-doubling, labeled by $e,f$ parity. \ce{H2O} photodissociation only excites the $\Omega = 1/2, f$ and $\Omega = 3/2, e$ states, which are usually labeled as the A' symmetry, with the remaining two states being labeled as the A'' symmetry. \\
\newline
We demonstrate in Fig. \ref{fig:OH_prompt} {the potential detection of} OH prompt emission right down to 9 $\mu$m in this source. These lines have been seen with JWST/MIRI in {a disk and} protostellar source \citep{zannese2024, neufeld2024}, where the asymmetry from the preferential excitation of the A' symmetry, uniquely indicative of OH production through \ce{H2O} photodissociation, was also demonstrated. Highly excited OH emission has been observed with \textit{Spitzer} in the DG Tau disk \citep{carr2014} from 10 $\mu$m onwards, and also more recently with MIRI in TW Hya \citep{henning2024}. \\
\newline
We present our {potential} detection of OH prompt emission in Fig. \ref{fig:OH_prompt} using a non-LTE OH emission model run with the thermochemical code Dust And LINes (DALI) from \citet{tabone2024} that includes the effects of prompt emission {produced by \ce{H2O} photodissociation}. The lines belonging to the A' symmetry are indicated in Fig. \ref{fig:OH_prompt} with thick, solid arrows, whereas the A'' lines are indicated with thin, dashed arrows. At short (<12 $\mu$m) wavelengths, only the {A' line pairs} are detected ({the individual lines} are spectrally unresolved from one another at this wavelength, but resolved from the two A'' lines), whereas the {A'' line pairs} are much weaker. This makes the set of lines appear as either a singlet ($\sim$9-10 $\mu$m) or a very asymmetric doublet ($\sim$10-12 $\mu$m). This asymmetry between the A' and A'' lines is uniquely indicative of prompt emission produced by \ce{H2O} photodissociation. We note that there are some fringe residuals in this wavelength region, which is a known issue. Still, our comparison to the depicted model provides a {compelling case for this} detection. \\
\newline
At wavelengths beyond $\sim$14 $\mu$m, it becomes clear that the component of the A'' symmetry is now also present. This indicates that, aside from \ce{H2O} photodissociation, chemical pumping {and collisional excitation} also provide an important component of the observed OH emission, as is expected for a disk \citep{mandell2008, salyk2008}. At this wavelength, the two A' lines and two A'' lines are almost spectrally resolved, forming a triplet. The DALI model now closely matches the best-fit LTE slab model (shown in blue in the insets in Fig. \ref{fig:OH_prompt}) which is purely symmetrical with an equal contribution from the A' and A'' symmetries. The DALI model still shows a very small asymmetry between the A' and A'' lines, indicating that the preferential population of the A' states by \ce{H2O} photodissociation affects not only the lines at shorter wavelengths, but also the triplets and quadruplets beyond 14 $\mu$m, though to a much smaller extent. {Given the potential detection of} OH prompt emission in our source, it stands to reason that this smaller asymmetry should be present in the emission we observe beyond 14 $\mu$m as well, but this effect is likely too small to observe. \\
\newline
We {find evidence of} OH prompt emission in CX Tau, despite the source not being a strong accretor. This could indicate some degree of dust growth and settling in the disk, allowing the UV (Ly$\alpha$) radiation from the star to reach deeper into the surface layers, photodissociating the \ce{H2O}. The shape of the silicate feature in CX Tau is consistent with this idea. We also note that the \ce{CO2} gas, which is typically located in a similar region \citep[e.g.][]{bosman2022a, bosman2022b}, would likely not be similarly affected, as its photodissociation rates in a radiation field dominated by Ly$\alpha$ have been found to be a few orders of magnitude below that of \ce{H2O} \citep{heays2017}.

\begin{figure*}
    \centering
    \includegraphics[width=0.8\textwidth]{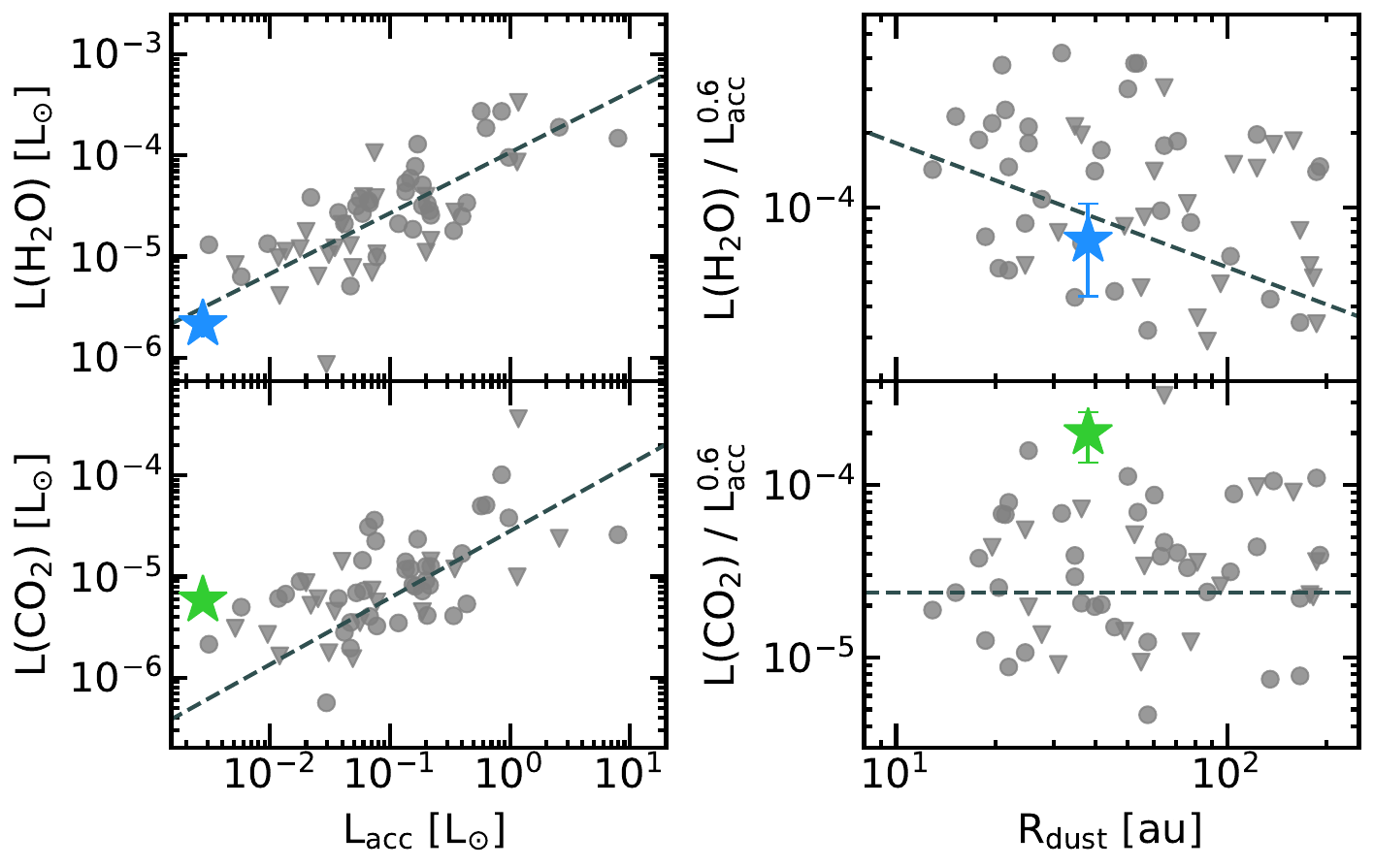}
    \caption{Comparison of the \ce{CO2} and \ce{H2O} luminosities derived for CX Tau (green and blue stars, respectively) to the data shown in \citet{banzatti2020} (shown in grey). {The left panels show the \ce{H2O} and \ce{CO2} luminosities as a function of accretion luminosity. The right panels show the \ce{H2O} and \ce{CO2} luminosities corrected for the correlation with accretion luminosity as a function of dust radius.} The reported correlation between the \ce{H2O} luminosity and the dust radius from \citet{banzatti2020} is shown in the top panel, and the median \ce{CO2} luminosity is shown in the bottom panel (both shown in a grey dashed line). {We note that the measurements from our data on both plots contain 1$\sigma$ error bars, but these do not exceed the marker size in the left panel.}}
    \label{fig:banzatti}
\end{figure*}

\section{Discussion}\label{sec:discussion}

\citet{banzatti2020} {find}, based on \textit{Spitzer} data, that compact dust disks typically have a higher \ce{H2O} line luminosity than more extended sources. This could be attributed to these disks efficiently transporting icy pebbles inwards and enriching their inner disks in \ce{H2O}. CX Tau is a very compact disk with one of the highest $R_{\rm gas}/R_{\rm dust}$ ratios at a value of 5 as measured by ALMA, indicating that radial drift is likely very efficient. However, our data show that the \ce{H2O} emission in CX Tau is not very visually prominent in the spectrum, compared with the forest of emission lines seen in some other compact disks, e.g. DR Tau \citep{temmink2024b, temmink2024a} and FZ Tau \citep{pontoppidan2024}. Instead, it contains a strong \ce{CO2} feature, allowing us to also detect emission from \ce{^13CO2} and potentially even \ce{CO^18O}. As such, it is interesting to explore how CX Tau fits into this picture of compact disks and radial drift.\\
\newline
We begin by comparing CX Tau to the sample of disks presented in \citet{banzatti2020}, depicted in Fig. \ref{fig:banzatti}. Using the same spectral windows to calculate the integrated fluxes \citep[see][]{najita2013}, we derive an \ce{H2O} flux of $(4.2 \pm 1.2) \times 10^{-15}$ erg s$^{-1}$ cm$^{-2}$ and a \ce{CO2} flux of $(11.5 \pm 1.5) \times 10^{-15}$ erg s$^{-1}$ cm$^{-2}$. We then convert these fluxes to a luminosity and plot them as a function of accretion luminosity, using the same value of $L_{\rm acc}$ for CX Tau as reported in \citet{banzatti2020} for consistency. This is depicted in the left panels of Fig. \ref{fig:banzatti}. From this, it becomes clear that both the \ce{H2O} and \ce{CO2} luminosities show a dependence on accretion luminosity. Our derived \ce{H2O} line strength falls within the expected range for a source with a low accretion luminosity, whereas the \ce{CO2} flux is stronger than expected from the derived correlation by almost an order of magnitude.\\
\newline
The right panels of Fig. \ref{fig:banzatti} show the relation between the \ce{H2O} and \ce{CO2} luminosities and the dust outer radius, {accounting for the correlation with $L_{\rm acc}$. We assume that this} value has an uncertainty of about 50\%, in line with typical uncertainties on the conversion factors from HI line flux to accretion luminosity. Here, we once again see that the \ce{H2O} line luminosity is in line with the rest of the \textit{Spitzer} sample, but the \ce{CO2} luminosity stands out, being brighter than the median \ce{CO2} flux by almost an order of magnitude. \\
\newline
As such, it seems that the \ce{H2O} flux of CX Tau falls in line with the \textit{Spitzer} sample from \citet{banzatti2020}, indicating that the low accretion rate provides a {likely} explanation for the weakness of the warm \ce{H2O} emission. For reference, both disks mentioned earlier, DR Tau and FZ Tau, have much higher accretion rates ($\sim$$10^{-7}\, M_\odot$ yr$^{-1}$; \citealt{mcclure2019, manara2023}) than CX Tau ($\sim$7 $\times 10^{-10}\, M_\odot$ yr$^{-1}$; \citealt{hartmann1998}). CX Tau also has a quite moderate inclination of 55\degree that could contribute to its weaker emission \citep[see, e.g.][]{banzatti2023_ground, banzatti2024}. However, while these two factors could explain the relative weakness of the \ce{H2O} emission, the \ce{CO2} emission seems to have a different story. It should be similarly affected, but instead seems stronger than expected. Thus, the question to explore becomes: why does CX Tau display such bright \ce{CO2} emission? We discuss several possibilities here.

\begin{figure}
    \centering
    \includegraphics[width=0.5\textwidth]{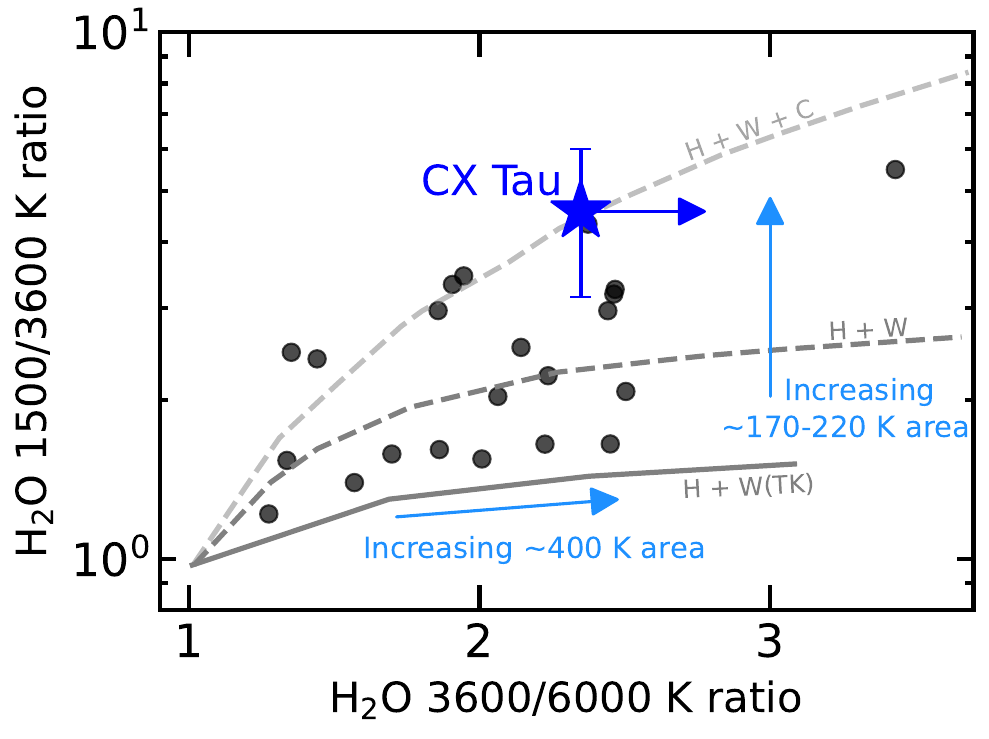}
    \caption{\ce{H2O} diagnostic diagram from \citet{banzatti2024}, adapted to include the measured line ratios for CX Tau. Data points from \citet{banzatti2024} (excluding data points with upper/lower limits) are indicated with black circles and the data point for CX Tau is indicated with a {blue} star. {Three} models from \citet{banzatti2024} are shown: {the model with a hot (850 K) and warm (400 K; optically thick) component is shown as a grey solid line (labeled ``H + W (TK)''), the model with a hot (850 K) and warm (400 K; optically thin) component is shown as a dashed, grey line (labeled ``H + W''), and the model with a hot (850 K), warm (400 K), and cold (190 K) component is shown as a dashed, light grey line (labeled ``H + W + C'').}}
    \label{fig:B24_fig9}
\end{figure}

\subsection{Radial drift}

{First}, the enhanced \ce{CO2} compared with \ce{H2O} could be caused by radial drift and the resulting chemical evolution of the disk. Modeling work by \citet{mah2023} has shown that the C/O ratio in the inner disk will decrease at early times due to the inwards drift of \ce{H2O}-ice-rich pebbles, whose contents sublimate and enrich the gas. Then, at later times, the C/O ratio will increase again due to this \ce{H2O}-rich gas draining onto the central star, and the inwards advection of C-rich gas from the outer disk. The work by \citet{kalyaan2021, kalyaan2023} shows also how the delivery of \ce{H2O} to the inner disk through pebble drift is sensitive to the location of potential gaps in the disk. The inner disk of CX Tau still seems to be in an O-rich phase, with \ce{H2O}, \ce{CO2} and OH firmly detected, but shifts in relative \ce{CO2 /H2O} abundances can already occur on a shorter timescale. Modeling work by \citet{sellek2024} shows that the enrichment in \ce{H2O} gas of the inner disk gas due to sublimation at its iceline will first enhance \ce{H2O} with respect to \ce{CO2}, as its iceline is closest to the star. As the \ce{H2O}-rich gas starts advecting inwards and subsequently drains onto the star, the gas that is relatively more enriched in \ce{CO2} near the \ce{CO2} iceline will also be advecting inwards, eventually enriching the inner disk in \ce{CO2} instead. \citet{mah2023} also showed that such a process proceeds faster for lower-mass stars. As CX Tau is rather low-mass (0.37 $M_\odot$), this could mean that it has already reached a more \ce{CO2}-rich phase in its evolution, earlier than higher-mass stars of a similar age. \\
\newline
\citet{sellek2024} also demonstrate that, while the inwards drift of \ce{H2O}-rich ice enriches the inner disk in \ce{H2O}, the associated influx of dust in the inner disk may obscure the additional emission from this gas. This is due to the assumption of a lower fragmentation velocity for the dry grains inside the \ce{H2O} snowline compared to icy grains outside the snowline \citep{blumwurm2008, gundlach2015}. As such, dust grains inside the \ce{H2O} snowline fragment more easily, thereby coupling to the gas and increasing the dust opacity in the inner disk (creating a ``traffic jam'', c.f. \citealt{pinilla2016}). {Based on these assumptions, these models predict that} the column density of \ce{H2O} retrieved from a synthetic spectrum {may be} completely insensitive to this inwards drift and enrichment of the inner disk, even though the total \ce{H2O} column density in this region is enhanced. The \ce{CO2} column density, however, is {predicted to be} much more sensitive to this evolution. This is caused by the fact that it is located further out in the disk and thus its emission is much less affected by the increase in opacity the dust drift causes, as this only affects the region inside the \ce{H2O} snowline. Therefore, \citet{sellek2024} find that the \ce{CO2} column density, and especially the \ce{CO2 /H2O} column density ratio, is a good tracer of how much of it has been brought to the inner disk.\\
\newline
Thus, it is possible that the bright \ce{CO2} emission in CX Tau is indicative of it currently undergoing a \ce{CO2}-rich phase. {If} the observed \ce{CO2} emission is enhanced due to the efficient drift, {it would explain} why it is stronger than expected from the source's low accretion luminosity (Fig. \ref{fig:banzatti}). The fact that the \ce{H2O} emission is not similarly enhanced could imply one of two scenarios: 1) any previous enhancement in \ce{H2O} vapour mass in the inner disk has already drained onto the star, and thus the observed \ce{H2O} flux is once again in line with the source's low accretion luminosity, or 2) the \ce{H2O} vapour mass in the inner disk is still enhanced due to the efficient radial drift (though past its peak), but this does not translate into an increase in observed \ce{H2O} flux due to the associated influx of dust. \\ 
\newline
Naturally, it is worth noting that \citet{sellek2024} also explore scenarios in which the sensitivity of \ce{H2O} emission to drift is increased. This is done, for example, by assuming the same fragmentation velocity for icy grains and dry grains, as more recent works suggest \citep{gundlach2018, musiolik2019}, which eliminates the effect of the traffic jam. Regardless, a \ce{CO2}-rich phase is still achieved. However, their models predict that the enhancement phases of \ce{H2O} and \ce{CO2} are not perfectly separated in time, and an enhanced \ce{CO2} vapour mass in the inner disk will likely coincide with an enhanced (though past its peak) \ce{H2O} vapour mass. As such, the previous scenario described above could provide a potential explanation for the absence of a bright forest of warm \ce{H2O} emission in CX Tau, where the enhancement from radial drift is hidden by the influx of dust. \\
\newline
There is certainly compelling evidence in the MIRI spectrum of CX Tau {indicating} that the radial drift of ices is important in this disk, as would be expected given its high $R_{\rm gas}/R_{\rm dust}$ ratio observed with ALMA \citep{facchini2019}. First, we report the detection of cold \ce{^13CO2} emission (Fig. \ref{fig:slab_13CO2}). Modeling work by \citet{bosman2017} has shown that the \ce{^13CO2} $Q$ branch feature is particularly sensitive to enhancements in the \ce{CO2} abundance near the iceline created by sublimating ices. A significant enhancement can make the feature stand out prominently among the surrounding \ce{^12CO2} $P$ branch lines, which is precisely what is observed in CX Tau. This, combined with the cold temperature the \ce{^13CO2} seems to be tracing, could indicate that there is indeed a significant volume of \ce{CO2} ice currently drifting across its iceline. The \ce{CO^18O} emission also shows an indication of a similarly cold temperature, so this emission may be an even better tracer of this phenomenon. \\
\newline
Second, we demonstrate the presence of a cold $\sim${200} K component in the observed \ce{H2O} emission (Fig. \ref{fig:slab_H2O_long}). The temperature of this component indicates that it potentially traces gas close to the snowline. \citet{banzatti2023} {and \citet{romero-mirza2024_sample}} demonstrate that this component is detected more strongly in the compact disks than the extended disks in their sample, which could indicate that radial drift of ices is responsible for enhancing this feature. This is further supported by the detection of a cold component in the compact disk DR Tau \citep{temmink2024b}. \\
\newline
{Additionally, this cold component has been further analyzed by \citet{banzatti2024} using ratios of several diagnostic \ce{H2O} lines (see Table 1 in that work) with $E_{\rm up}$ of 1500, 3600 and 6000 K. These line ratios provide useful information on the relative emission from different temperature components present in the spectrum. The 3600/6000 K line ratio mainly captures the strength of a warm ($\sim$400 K) component, whereas the 1500/3600 K line ratio captures the strength of a cold ($\sim${170-220} K) component. We measure these line ratios for CX Tau and demonstrate our results in Fig. \ref{fig:B24_fig9}. The 6000 K line is only marginally detected, so a 1$\sigma$ upper limit is reported. The measured line ratios closely match the model from \citet{banzatti2024} with a prominent cold, $\sim${190} K component (see their Sect. 4.1 and 4.2). This demonstrates that the \ce{H2O} emission in CX Tau has a significant contribution from a cold component that likely originates near the \ce{H2O} snowline.}\\
\newline
{Thus, the anomalously strong \ce{CO2} emission, along with the (potential) detection of two of its weaker isotopes, could be caused by the strong radial drift in this disk creating a \ce{CO2}-rich chemistry in its inner disk. The \ce{H2O} emission is visually much weaker. This seems contradictory for a drift-dominated disk, however, a cold \ce{H2O} component is still clearly detected, which is indicative of radial drift. The weakness of the warm \ce{H2O} emission can likely be explained by its low accretion luminosity. Additionally, even if the strong radial drift has enhanced both the \ce{CO2} and \ce{H2O} vapour mass in the inner disk, the associated influx of dust could potentially explain why an enhancement in observed flux is only seen in \ce{CO2}, and not in \ce{H2O}. The relatively weak \ce{C2H2} and HCN emission from the disk is also consistent with this scenario, as these species are likely more sensitive to CX Tau's low accretion luminosity than a process like radial drift.}

\subsection{A small inner cavity}

{An alternate explanation for the bright \ce{CO2} emission that has been proposed in previous work is} the presence of a small inner cavity. This argument was put forth to {potentially} explain the bright \ce{CO2} observed in GW Lup \citep{grant2023}, and it was tested with Dust And LINes (DALI) thermo-chemical models in \citet{vlasblom2024}. {The latter} work shows that the presence of a small, inner cavity can suppress the \ce{H2O} emission and enhance that of \ce{CO2}. {The work also} shows that a source like CX Tau, with a luminosity of $L_* = 0.2 L_\odot$, would need a cavity of approximately 2 au in radius (thus 4 au in diameter; see Appendix B in that work). This is below the resolution of the current highest angular resolution ALMA observations available \citep[][with a resolution of 5 au]{facchini2019}, and thus the presence of such a cavity cannot currently be confirmed or ruled out. \citet{facchini2019} do calculate an upper limit on the size of a possible inner cavity from the intensity profile and find an upper limit of 0.54 au. That would seem to rule out our proposed cavity of 2 au, but we argue that a 2 au cavity could still be completely hidden from view in these observations, especially if the cavity is not fully depleted in dust {(\citealt{vlasblom2024} show that a depletion in gas and dust of at least a factor $10^4$ is needed)}. Thus, even higher angular resolution ALMA observations, combined with super-resolution techniques, would be needed to truly confirm or rule out the presence of an inner cavity.\\
\newline
The presence of such a cavity does raise quite a few concerns when combined with our findings in this work. For one, we find very small emitting radii for most of our species, {on the order of $\sim$0.05} au. In Sect. \ref{subsec:slab_models} we stress that the emitting area is parameterized by a radius, but that the emission can also originate from a thin annulus further out. 
We can contrast this with the 2D thermochemical models presented in \citet{vlasblom2024}, which do show that the emission will originate from a very thin annulus once the cavity {is made large enough} to enhance the \ce{CO2} emission {sufficiently} with respect to the \ce{H2O} emission (see, e.g. Fig. 2 and Appendices A and B in that work), {though perhaps not quite to the extent our derived emitting area would imply.} 
\\
\newline
Similarly, one may wonder if such a cavity around a relatively faint star could even produce temperatures at the cavity wall that are hot enough to match our observed emission {of $\sim$500 K}. The thermochemical model by \citet{vlasblom2024} finds that temperatures of 500-600 K are reached within the molecular emitting region at the cavity wall, so this seems to be possible. We also refer back to CX Tau's moderate inclination of 55$\degree$ here. With this, the cavity wall would be directly exposed to the observer, so the high temperatures at the very inner edge of the cavity wall would be directly visible. \\ 
\newline
Finally, one should also consider that a 2 au cavity in the inner disk will likely impact the SED quite strongly. CX Tau actually has an interesting history of having once been classified as a transition disk based on the (lack of) near-IR excess in its SED \citep{najita2007}. However, when common color criteria are used, such as in \citet{furlan2011}, CX Tau does not qualify. {Theoretical work shows that IR spectral indices $n_{13-30}>0$ are likely a result of dust cavities \citep[e.g.][]{woitke2016, ballering2019}. \citet{banzatti2020} report an $n_{13-30}$ index of -0.15 for CX Tau. Nevertheless, their sample contains some disks with $n_{13-30} < 0$ that still have an inner cavity detected ({which, interestingly, are sources with similar inclinations to CX Tau, as seen by the red markers in, e.g., their Figs. 6 and 9}), indicating that the SED of CX Tau does seem to allow for the presence of a small cavity.} \\
\newline
{To conclude, the presence of a small, inner cavity in CX Tau cannot yet be ruled out. However, the scenario does raise some valid concerns, as modeling has found that the needed cavity size (2 au) is relatively large, given CX Tau's small dust radius and low luminosity. Thus, while a small cavity could be responsible for the enhanced \ce{CO2} emission, the scenario of enhancement by radial drift seems preferred.}

\subsection{Gas-to-dust ratio and disk temperature}

{Finally, \citet{vlasblom2024} show that a large amount of dust in the inner regions could also enhance \ce{CO2} emission relative to \ce{H2O}, as shown from their models with a lower gas-to-dust ratio.} This dust could be brought to the inner disk by radial drift, and subsequently could be stirred up into the IR emitting layers of the disk by turbulence, clouding the region. It has been shown that the \ce{H2O} emission detected in disks with \textit{Spitzer}, which are thus disks with much brighter \ce{H2O} emission than CX Tau, is likely coming from a layer with a gas-to-dust ratio {that is locally enhanced} by 1-2 orders of magnitude \citep{meijerink2009, bosman2022a}. An enhancement in the amount of dust in these layers could thus obscure this emission, as is also seen in \citet{woitke2018} and \citet{greenwood2019}. Furthermore, \citet{bosman2023} speculate that the features from \ce{CO2} could remain visible due to IR pumping, though in-depth modeling would be needed to confirm this. 
The dust would also make the IR emitting layers cooler. As \ce{CO2} preferentially forms from OH over \ce{H2O} at temperatures below $\sim$300 K \citep{charnley1997, vandishoeck2013, walsh2015}, this means that more \ce{CO2} could be formed in a cooler disk with more dust than in a hotter, more settled disk. \\
\newline
{This scenario is proposed by \citet{bosman2023} to explain the bright \ce{CO2} emission observed in IM Lup, another source that seems to have peculiarly bright \ce{CO2} emission, just like GW Lup and CX Tau. \citet{bosman2021, bosman2023} demonstrate a pile-up of dust in the inner disk of IM Lup, pointing to a low gas-to-dust ratio and high turbulence in this region. This scenario may thus provide another alternate explanation for the bright \ce{CO2} emission in CX Tau, though one would need to investigate further whether the source shows similar evidence for a low gas-to-dust ratio in the inner disk. }

\section{Conclusions}\label{sec:conclusions}

We present JWST MIRI/MRS observations of the disk around CX Tau, a compact, drift-dominated disk whose IR molecular features have not been analyzed before. Our main conclusions are as follows:

\begin{itemize}
    \item We detect bright emission from \ce{CO2} and much weaker features from \ce{H2O, ^13CO2, C2H2, HCN}, and OH for the first time in this disk. We constrain their properties using 0D LTE slab models. All of these weaker features have line-to-continuum ratios that are too small for them to have been detected previously, demonstrating the vast improvement in the detection of such faint emission that JWST has brought.
    \item We find the \ce{^12CO2} to be optically thick, tracing a temperature of $\sim${450} K {at an (equivalent) emitting radius of $\sim$0.05 au}, whereas the \ce{^13CO2} traces much colder temperatures ($\sim${200} K) and {a larger emitting area}.  
    \item {We also report a potential detection of the even rarer isotopologue, \ce{CO^18O}, in the disk.}
    \item {We detect warm, $\sim$500 K, pure rotational emission from \ce{H2O}, as well as ro-vibrational emission that traces slightly warmer gas. We also find evidence for a colder, $\sim$200 K, component.}
    \item We {report a potential detection of} highly excited rotational OH lines between 9 and 12 $\mu$m which are caused by \ce{H2O} photodissociation, as well as emission caused by chemical pumping {and collisional excitation} at longer wavelengths. 
    \item We detect 4 pure rotational \ce{H2} lines for which we find tentative evidence that this emission is extended, based on analysis of the PSF.
    \item {The strong radial drift, known to be present in this source, could be responsible for the observed \ce{CO2}-rich chemistry. The drift could have increased the \ce{CO2} vapour mass in the inner disk, enhancing the observed emission. As it is located further out in the disk than \ce{H2O}, the \ce{CO2} emission would be less affected by any enhancements in dust opacity from the radial drift.}
    \item {The comparatively weaker \ce{H2O} emission could be explained by the source's low accretion luminosity, assuming that any enhancement in \ce{H2O} vapour mass by drift has already advected onto the star. Alternatively, the \ce{H2O} vapour mass could still be somewhat enhanced in the inner disk, but it is possible that this would not translate into an increased flux due to the increased dust opacity within the \ce{H2O} snowline.}
    \item The observed cold \ce{^13CO2} and \ce{H2O} support the idea that radial drift of ices is important for the observed IR emission in this disk.
    \item {Alternatively,} the bright \ce{CO2} emission and relatively weaker \ce{H2O} emission could hint at the presence of previously unrevealed substructures in the disk in the form of a small, inner cavity with a size of roughly 2 au in radius. Higher angular resolution ALMA observations are needed to confirm this.
\end{itemize}
This work has demonstrated how JWST can give us a much deeper insight into how disk structure and radial drift of ices could produce the \ce{CO2}-rich chemistry observed in this drift-dominated disk {on scales where terrestrial planets may be forming}. To understand these processes and their link to disk chemistry even better, it will be interesting to investigate these trends in a larger sample, for example in a sample of compact, drift-dominated disks (e.g. expanding upon the work by \citealt{banzatti2023}), or in a sample of \ce{CO2}-bright disks.

\begin{acknowledgements}

{We thank the referee for their detailed comments which helped improve this paper.}

This work is based on observations made with the NASA/ESA/CSA James Webb Space Telescope. The data were obtained from the Mikulski Archive for Space Telescopes at the Space Telescope Science Institute, which is operated by the Association of Universities for Research in Astronomy, Inc., under NASA contract NAS 5-03127 for JWST. These observations are associated with program \#1282. The following National and International Funding Agencies funded and supported the MIRI development: NASA; ESA; Belgian Science Policy Office (BELSPO); Centre Nationale d’Etudes Spatiales (CNES); Danish National Space Centre; Deutsches Zentrum fur Luft- und Raumfahrt (DLR); Enterprise Ireland; Ministerio de Econom\'ia y Competividad; Netherlands Research School for Astronomy (NOVA); Netherlands Organisation for Scientific Research (NWO); Science and Technology Facilities Council; Swiss Space Office; Swedish National Space Agency; and UK Space Agency.

M.V., M.T. and ADS acknowledge support from the ERC grant 101019751 MOLDISK.
A.C.G. acknowledges from PRIN-MUR 2022 20228JPA3A “The path to star and planet formation in the JWST era (PATH)” funded by NextGeneration EU and by INAF-GoG 2022 “NIR-dark Accretion Outbursts in Massive Young stellar objects (NAOMY)” and Large Grant INAF 2022 “YSOs Outflows, Disks and Accretion: towards a global framework for the evolution of planet forming systems (YODA)”.
D.B. is funded by the Spanish MCIN/AEI/10.13039/501100011033 grant PID2019-107061GB-C61.
G.P. gratefully acknowledges support from the Max Planck Society.
E.v.D. acknowledges support from the ERC grant 101019751 MOLDISK and the Danish National Research Foundation through the Center of Excellence ``InterCat'' (DNRF150). 
T.H. and K.S. acknowledge support from the European Research Council under the Horizon 2020 Framework Program via the ERC Advanced Grant Origins 83 24 28. 
I.K., A.M.A., and E.v.D. acknowledge support from grant TOP-1 614.001.751 from the Dutch Research Council (NWO).
I.K. acknowledges funding from H2020-MSCA-ITN-2019, grant no. 860470 (CHAMELEON).
B.T. is a Laureate of the Paris Region fellowship program, which is supported by the Ile-de-France Region and has received funding under the Horizon 2020 innovation framework program and Marie Sklodowska-Curie grant agreement No. 945298.
V.C. acknowledges funding from the Belgian F.R.S.-FNRS.
D.G. thanks the Belgian Federal Science Policy Office (BELSPO) for the provision of financial support in the framework of the PRODEX Programme of the European Space Agency (ESA).

\end{acknowledgements}

%
%
\bibliographystyle{aa}
\bibliography{references}

\begin{appendix}
\section{Slab fitting routine}\label{app:slab}

We show here several additional figures from our slab fitting routine. In Figs. \ref{fig:slab_15um_ind_p1}, \ref{fig:slab_15um_ind_p2} and \ref{fig:slab_15um_ind_closeups}, we show the six individual fits performed in the 13.5 to 17.5 $\mu$m region, and we discuss the {uncertainties on the fitting routine} below. In Fig. \ref{fig:chi2_15um} we show the $\chi^2$ plots that result {from the \ce{CO2, ^13CO2, C2H2} and \ce{H2O} fits (as the OH and HCN fits are found to be unconstrained).} In Fig. \ref{fig:residuals_15} we discuss the residuals in the fit which potentially belongs to emission from \ce{CO^18O}. 

\subsection{{Uncertainties on the step-by-step fitting routine}}\label{app:fit}

{As stated in Sect. \ref{subsec:molecular}, the retrieved fit parameters are quite uncertain due to the relatively low S/N of the spectrum and the weakness of the molecular emission. In the 13.5-17.5 $\mu$m range, weak features from \ce{^12CO2, ^13CO2, H2O, C2H2, HCN}, and OH all overlap, making the emission complicated to fit. We do this by fitting the emission of each molecule individually, after which it is subtracted from the spectrum and the next species is fit. It is also possible to fit the emission from all species simultaneously using an MCMC routine \citep[e.g., as done in][]{grant2024}. We provide a comparison between the two methods in Fig. \ref{fig:MCMC}, which demonstrates that the two produce similar conclusions. We also stress that our main results do not rely on the emission properties being constrained with large accuracy.} \\
\newline
As demonstrated in Figs. \ref{fig:slab_15um_ind_p1} and \ref{fig:slab_15um_ind_p2}, we begin by fitting the most prominent feature: \ce{^12CO2}. After that, we fit the \ce{H2O} emission, of which the features span the entire fitted wavelength range. After that, we fit the remaining molecular features from \ce{C2H2}, \ce{^13CO2}, OH, and HCN  in that order. {The selection of the windows in which the fits are performed can significantly impact the results of this method. After all, a lot of the weak molecular features have significant overlap with other species. As such, we are careful to select features that are as isolated as possible from contaminating emission. This is particularly important for \ce{^12CO2} and \ce{H2O}.} The remaining species have detected features only in more specific wavelength ranges, {thus strongly limiting the possible isolated features that can be selected. Thus, we focus on avoiding their features in the \ce{^12CO2} and \ce{H2O} fit windows. \\
\newline
For \ce{H2O}, we find that the retrieved column density is particularly sensitive to the selection of the fit windows: enlarging the windows allows for more weak features at the noise level to be included, increasing the column density. Additionally, it is important that our fit windows} contain lines of different upper level energies $E_{\rm up}$ and Einstein A-coefficients $A_{\rm ij}$, to avoid a potential bias in the fit. This is especially important for \ce{H2O}, for which lines with a large diversity in $E_{\rm up}$ and $A_{\rm ij}$ are present in this wavelength range \citep[see, e.g., Figs. 14 and 15 in][]{pontoppidan2024}.\\
\newline
{As such, we define our fit windows as follows: for \ce{^12CO2}, our windows contain its $Q$ branch, several isolated $P$ and $R$ branch lines, as well as the hot bands. For \ce{H2O}, we select the detected, unblended lines from Tables 6 \& 7 in \citet{banzatti2024} in addition to a few other clearly detected features, e.g. at 17.2 $\mu$m. The remaining species have well-detected features only in more specific wavelength ranges. For OH, we make sure to cover all triplets seen in the 13.5-17.5 $\mu$m range, avoiding lines blended with \ce{H2O}. For \ce{^13CO2, C2H2} and HCN we cover their entire $Q$ branches in our windows, but we do not cover much more as these are the only prominent features we can detect, aside from the HCN hot band at 14.3 $\mu$m. The windows are indicated in Figs. \ref{fig:slab_15um_ind_p1} and \ref{fig:slab_15um_ind_p2} with horizontal bars. }\\
\newline
Finally, the specific order in which these six species are fit may also impact the fit results. {The features from \ce{C2H2}, \ce{^13CO2}, OH, and HCN do not overlap with one another significantly, so the order in which they are fit does not have a significant impact. We do find, however, that it is important to remove the contribution from \ce{^12CO2} emission before the \ce{H2O}, as the latter is much weaker and thus blended with the \ce{^12CO2} $P$ and $R$ branch lines.} We also find that the \ce{H2O} emission is significantly blended with the \ce{C2H2} and HCN features, making it important to fit the \ce{H2O} before those two species. {However, selecting isolated fit windows for each species can mitigate this, as evidenced by the fact that the MCMC method, which fits all species simultaneously, produces similar conclusions as the step-by-step method.} \\
\newline
{The best-fit parameters of the \ce{^12CO2, ^13CO2, C2H2}, and \ce{H2O} fits and any degeneracies between them can be seen in Fig. \ref{fig:chi2_15um}. We exclude the $\chi^2$ maps from OH, HCN, and \ce{H2O} at 5.5-8.5 $\mu$m. The HCN and OH fits are unconstrained and therefore do not provide a reliable indication of the emission properties. The \ce{H2O} fit at short wavelengths (as well as the OH fit) may be affected by non-LTE effects and the retrieved column density is likely overestimated. For the other species, the parameters are more constrained. However, the uncertainties on these values are still large and thus we do not report on the best-fitting parameters in much detail. \ce{^12CO2} is the only exception to this, being the brightest emission feature. {Still, we do caution that the hot band at 16.2 $\mu$m is affected by an artifact, which may have affected the estimation of our fit results slightly.} For the other species, we can only derive the temperature within $\sim$100-200 K and the column density within $\sim$1-2 dex. } 

\subsection{{Comparison with the MCMC routine}}

{We fit the emission from \ce{^12CO2, ^13CO2, H2O, C2H2, HCN}, and OH between 13.5 and 17.5 $\mu$m simultaneously using an MCMC routine, following the method described in \citet{grant2024}. We compare the best-fit results from this method to the best fits from our step-by-step routine in Fig. \ref{fig:MCMC}. The conclusions derived from both methods generally agree, suggesting that the step-by-step fitting is not significantly biased by the order in which we fit the molecules. The \ce{^12CO2} parameters are the most well-constrained. For \ce{H2O}, the temperature and emitting area agree well, but the derived column density from the MCMC is higher (> 10$^{20}$ cm$^{-2}$). This is likely due to the inclusion of weak features at the noise level, as we performed the fit over the whole wavelength range at once rather than limiting it to specific spectral features. As such, the MCMC method may be better suited for fitting the \ce{H2O} emission in higher S/N sources rather than a low S/N source like CX Tau.} \\
\newline
{The MCMC finds both the \ce{C2H2} and \ce{^13CO2} emission to be optically thin and thus the column density and emitting area are completely degenerate. For \ce{C2H2}, this agrees with our step-by-step fitting. We stress that the MCMC was allowed to explore a broader parameter space and the slight difference between our results and the MCMC is due to this degeneracy. For \ce{^13CO2}, the MCMC seems to favor a more optically thin solution than our step-by-step routine, though this parameter is poorly constrained. The cold temperature of the \ce{^13CO2} emission, however, is agreed upon by both methods.}  

\subsection{Residuals}\label{app:residuals}

\begin{figure*}
    \centering
    \makebox[\textwidth][c]{\includegraphics[width=1.\textwidth]{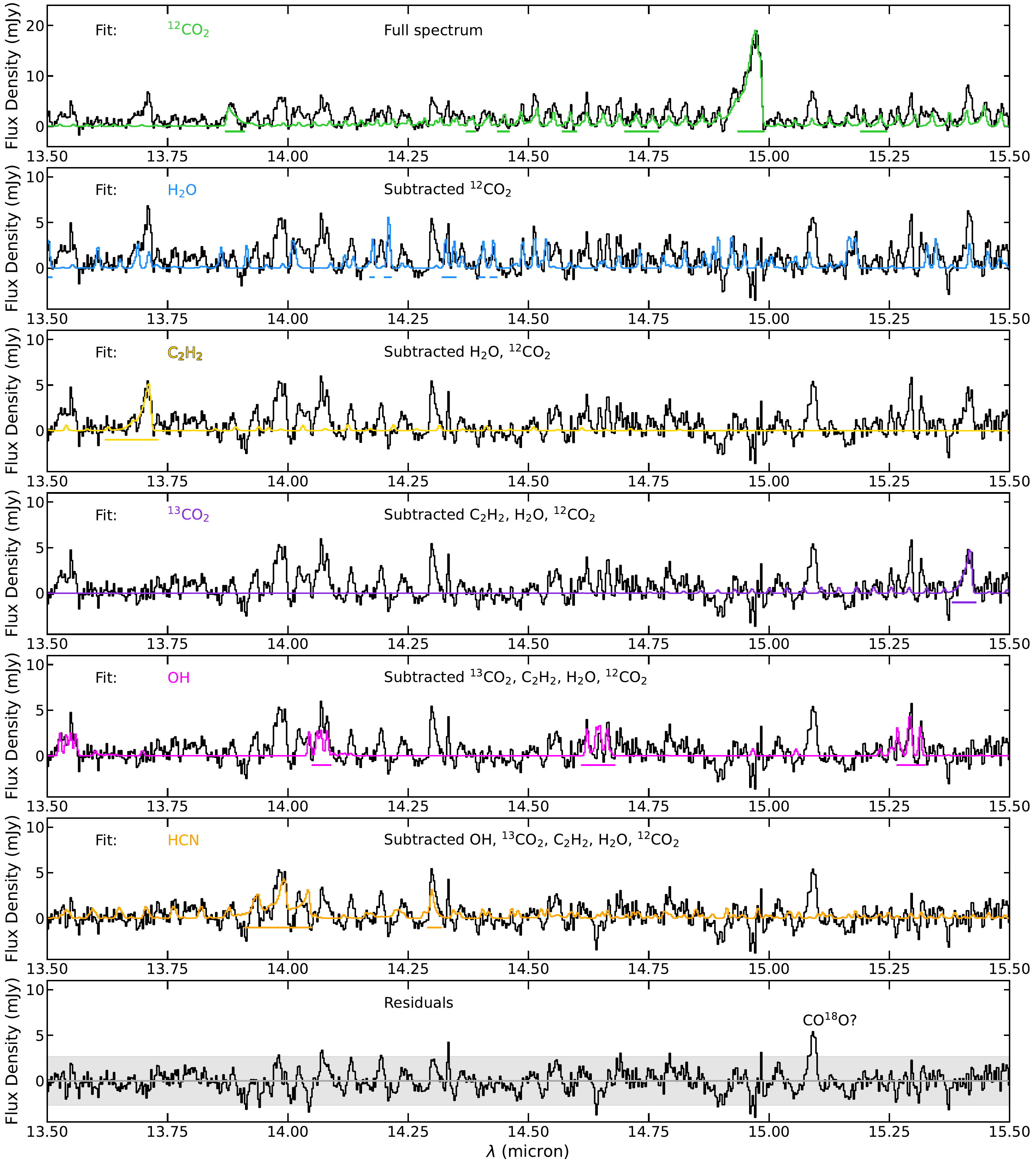}}
    \caption{A zoom-in of the {13.5-15.5} $\mu$m region of the CX Tau spectrum (black), where all individual slab model fits are shown. Horizontal bars indicate {which spectral windows were used} for the $\chi^2$ fit, and each panel indicates which other models were removed from the spectrum before the fit was performed, to avoid contamination from other molecules. {The final panel shows the residuals after all fits are subtracted, with the shaded region indicating that most residuals fall below 3$\sigma$. We note the difference in scale on the y axis between the top panel and the rest.} }
    \label{fig:slab_15um_ind_p1}
\end{figure*}

\begin{figure*}
    \centering
    \makebox[\textwidth][c]{\includegraphics[width=1.\textwidth]{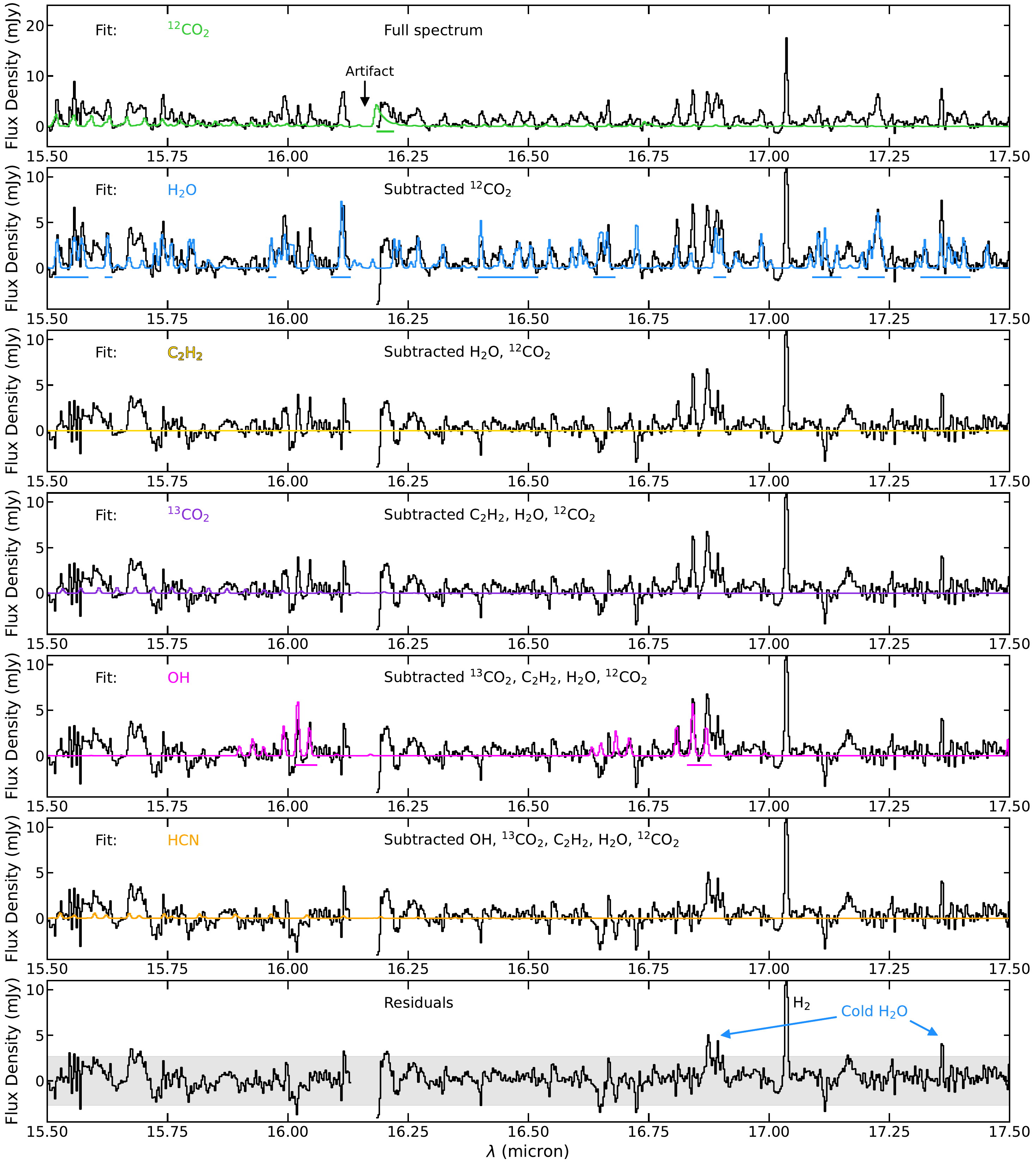}}
    \caption{A zoom-in of the {15.5-17.5} $\mu$m region of the CX Tau spectrum (black), where all individual slab model fits are shown. Horizontal bars indicate {which spectral windows were used} for the $\chi^2$ fit, and each panel indicates which other models were removed from the spectrum before the fit was performed, to avoid contamination from other molecules. {The final panel shows the residuals after all fits are subtracted, with the shaded region indicating that most residuals fall below 3$\sigma$. {An artifact around 16.15 $\mu$m has been masked in all panels.} We {also} note the difference in scale on the y axis between the top panel and the rest.}}
    \label{fig:slab_15um_ind_p2}
\end{figure*}

\begin{figure*}
    \centering
    \makebox[\textwidth][c]{\includegraphics[width=0.85\textwidth]{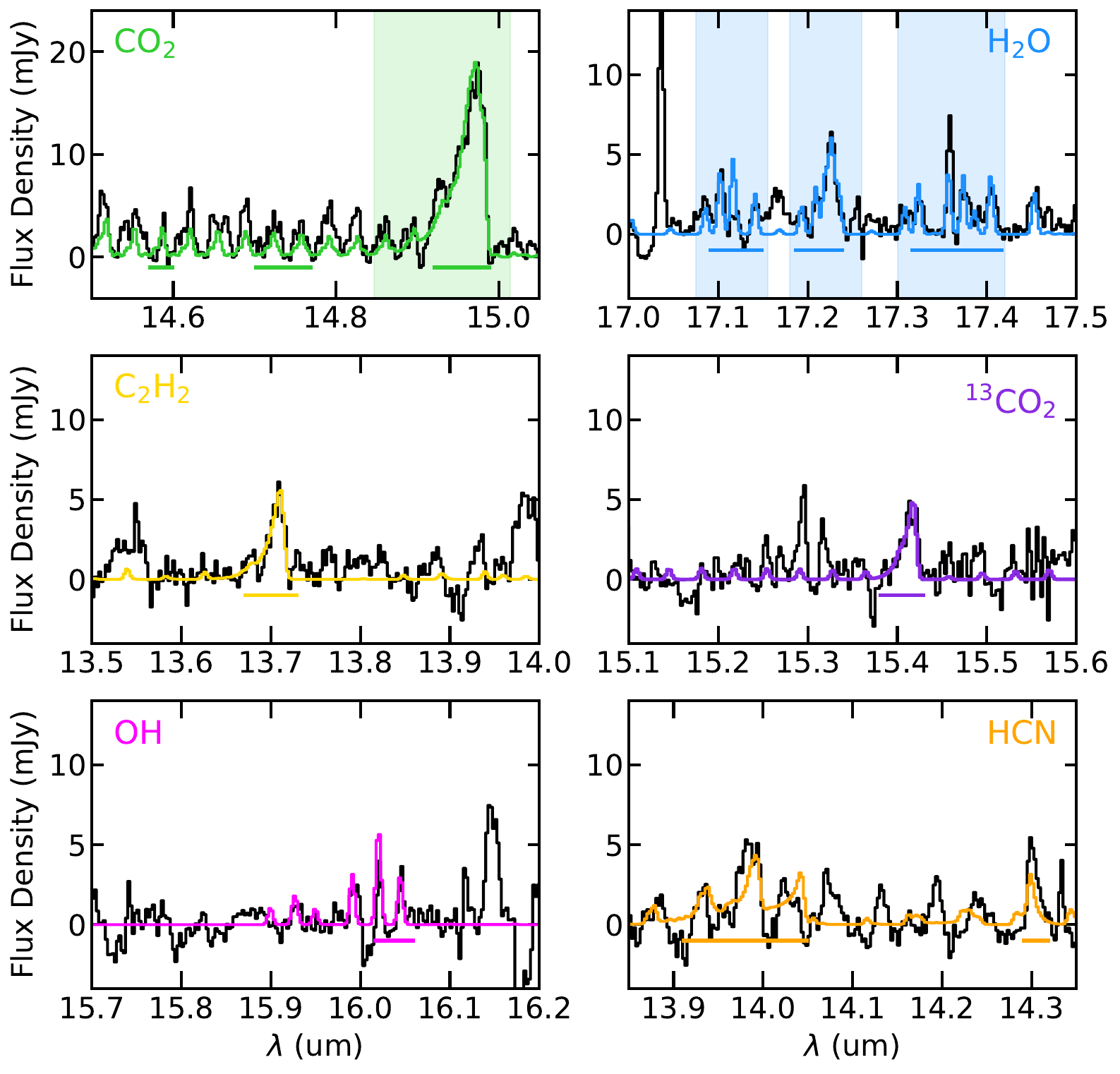}}
    \caption{{Various zoom-ins of Figs. \ref{fig:slab_15um_ind_p1} and \ref{fig:slab_15um_ind_p2}, to depict the goodness of the individual slab fits. Horizontal bars indicate {which spectral windows were used} for the $\chi^2$ fit. The vertical shaded regions indicate the windows used to calculate the \ce{CO2} and \ce{H2O} fluxes, following \citet{salyk2011} and \citet{najita2013}.}}
    \label{fig:slab_15um_ind_closeups}
\end{figure*}

\begin{figure*}
    \centering
    \includegraphics[width=0.95\textwidth]{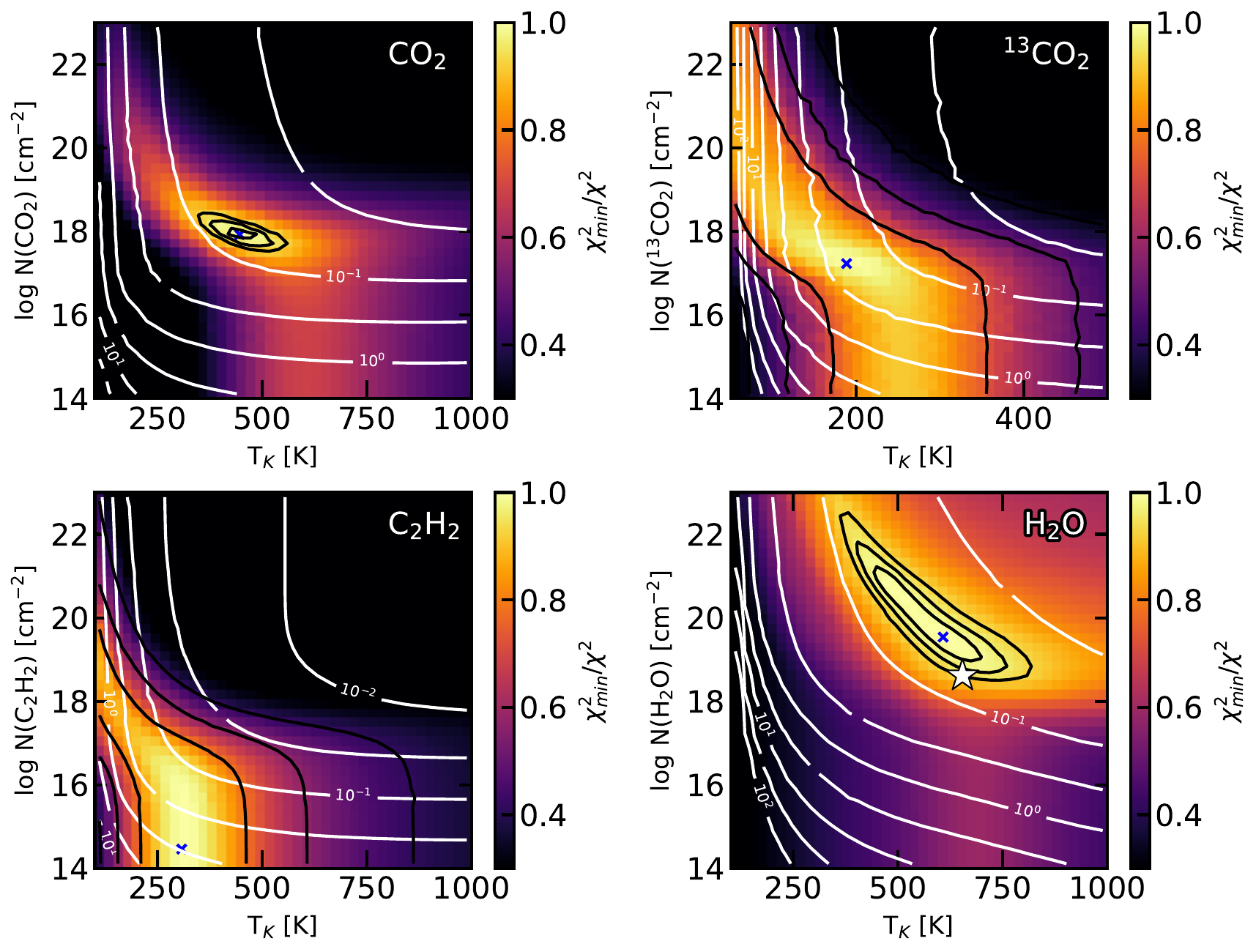}
    \caption{$\chi^2$ maps for the \ce{^12CO2, ^13CO2, C2H2} and \ce{H2O} slab model fits in the 13.5-17.5 $\mu$m wavelength range. The color-scale indicates $\chi_{\rm min}^2/\chi^2$, the three black contours correspond to the 1$\sigma$, 2$\sigma$, and 3$\sigma$ levels, and the white contours show the emitting radii in au. The best-fit model is indicated with a blue cross, and corresponds to $\chi_{\rm min}^2/\chi^2=1$. On the bottom-right panel, the best \ce{H2O} fit using the isolated lines from Tables 6 and 7 from \citet{banzatti2024} is depicted with a white star.}
    \label{fig:chi2_15um}
\end{figure*}

\begin{figure*}
    \centering
    \includegraphics[width=0.95\textwidth]{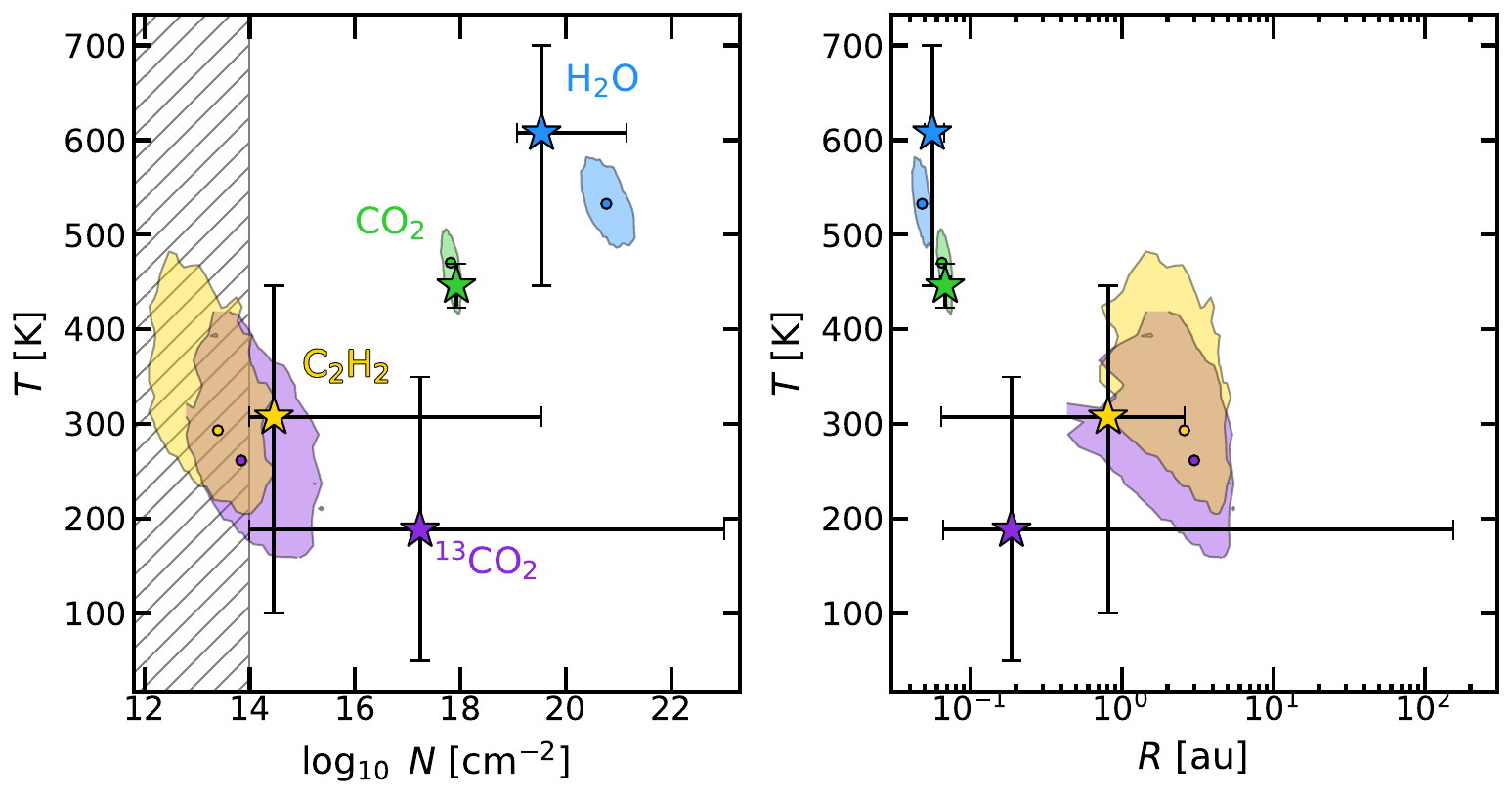}
    \caption{{Comparison between the MCMC routine and our step-by-step fitting routine. The stars represent the best-fit results from the step-by-step fits with 1 $\sigma$ errorbars. The small circles represent the best-fit results from the MCMC and the colored area represents a 3 $\sigma$ confidence. The hatched area in the left panel demonstrates the parameter space that is covered by the MCMC, but not the step-by-step routine.} }
    \label{fig:MCMC}
\end{figure*}

\begin{figure*}
    \centering
    \makebox[\textwidth][c]{\includegraphics[width=\textwidth]{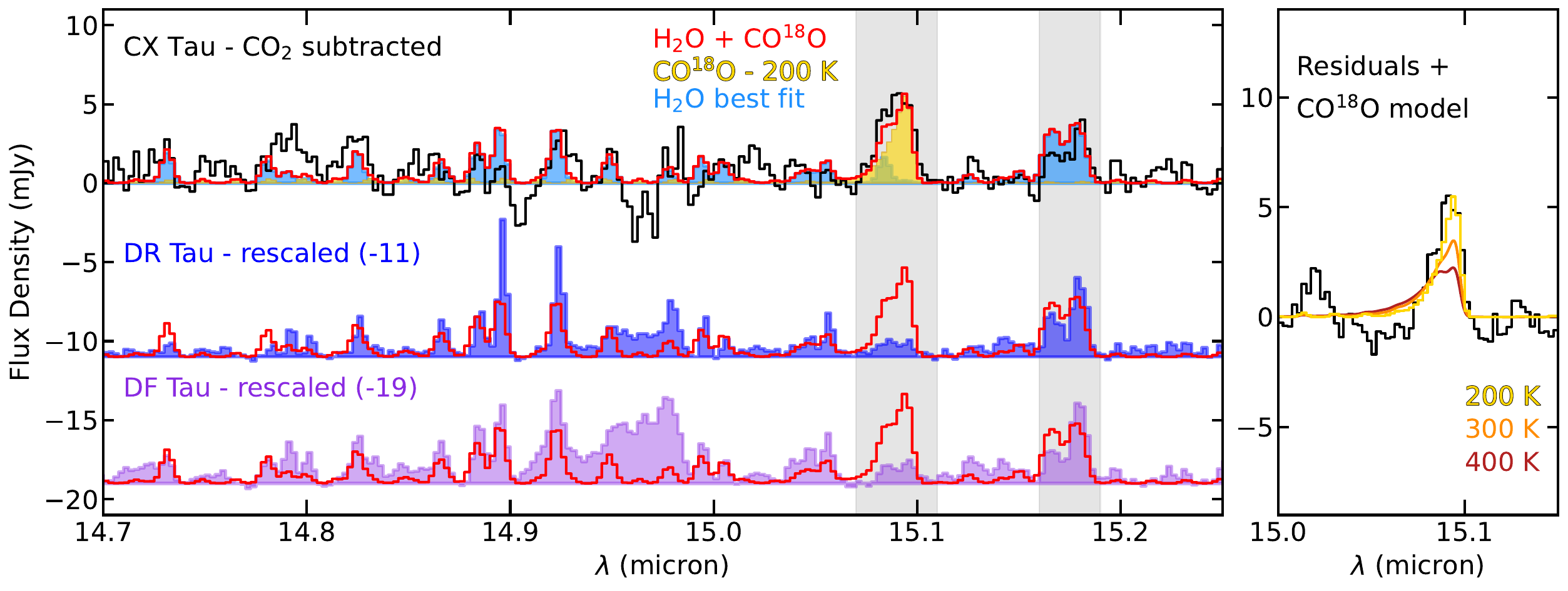}}
    \caption{{A zoom-in of the CX Tau spectrum around the 15.07 $\mu$m feature which stands out in our residuals in Fig. \ref{fig:slab_15um_ind_p1}. Left: The CX Tau spectrum with the best-fit \ce{CO2} model subtracted is shown in black, the best-fit \ce{H2O} model is shown in light blue, a 200 K, $N = 10^{16}$ cm$^{-2}$ \ce{CO^18O} model shown in yellow and the combined \ce{H2O} and \ce{CO^18O} model is shown in red. The spectra of DR Tau and DF Tau \citep{temmink2024b, temmink2024a, grant2024}, rescaled to the 15.17 $\mu$m \ce{H2O} feature in CX Tau, are shown in dark blue and purple, respectively. Right: The CX Tau spectrum with all best-fit models (\ce{^12CO2, ^13CO2, H2O, C2H2, HCN} and OH) subtracted is shown in black. Three \ce{CO^18O} models with $N = 10^{16}$ cm$^{-2}$ and different temperatures are shown in yellow, orange and brown. } }
    \label{fig:residuals_15}
\end{figure*}

We demonstrate in Figs. \ref{fig:slab_15um_ind_p1} and \ref{fig:slab_15um_ind_p2} that most of the residuals in the 13.5-17.5 $\mu$m region fall within 3$\sigma$ after all of the fits are subtracted. The residuals at the shorter wavelengths, between 13.5 and 16 $\mu$m, are likely affected by the heavy blending of weak features from \ce{CO2, C2H2, HCN, H2O}, and OH, perhaps explaining their observed structure. It is also possible that some weak \ce{H2O} features from hotter emission are unaccounted for in our single-temperature slab fit. In fact, the features depicted in Fig. \ref{fig:slab_15um_ind_p2} at 16.9 and 17.35 $\mu$m coincide with low $E_{\rm up}$ \ce{H2O} lines, and can thus likely be attributed to a colder component of \ce{H2O}, indicating that there are multiple temperature components present in this wavelength range. However, as also mentioned in Sect. \ref{subsub:H2O}, we do not intend to characterize the emission properties in that much detail in this work, and as such we only present a single-temperature slab fit. \\
\newline
{Aside from that, one residual feature stands out beyond 3$\sigma$: a peak at 15.07 $\mu$m.} This feature is of particular interest, as it coincides with the $Q$ branch of \ce{CO^18O}. We aim to explore the possibility of a detection with Fig. \ref{fig:residuals_15}. Here, we compare our data for CX Tau to two \ce{H2O}-rich disks, DR Tau \citep{temmink2024b, temmink2024a} and DF Tau \citep{grant2024}. These disks contain evidence for hotter and colder water components that are not included within our single-component fit. We rescale the spectra for these two disks to the \ce{H2O} feature at 15.17 $\mu$m. In the left panel of the figure, we show the CX Tau spectrum with the \ce{CO2} emission subtracted in black, the best-fit \ce{H2O} model in light blue, a 200 K \ce{CO^18O} model in yellow, and the combination of the two models (\ce{H2O} and \ce{CO^18O}) in red. We compare this to the two \ce{H2O}-rich sources, which demonstrates that the 15.07 $\mu$m feature seems to stand out abnormally in CX Tau. A comparison to the neighboring 15.17 $\mu$m \ce{H2O} feature makes this especially clear. \\
\newline
DR Tau and DF Tau both show \ce{H2O} emission at these two wavelengths, indicating that at least some of the 15.07 $\mu$m feature can be ascribed to \ce{H2O}. It is possible that our current best fit lacks some high-energy \ce{H2O} emission, as it seems too narrow compared to the emission seen in DR Tau and DF Tau. However, it is also clear that the 15.17 $\mu$m lines are much stronger than the 15.07 $\mu$m lines in DR Tau and DF Tau, whereas the opposite is true in CX Tau. Thus, if one tries to fit a high temperature \ce{H2O} model that fills in the 15.07 $\mu$m feature to the extent that is seen in CX Tau, the surrounding \ce{H2O} lines will be massively overproduced. This lends credit to the idea that at least some of the prominence of this feature is due to a contribution of \ce{CO^18O} emission, not solely \ce{H2O}. \\
\newline
This is supported by the \ce{CO^18O} models shown in the right panel of Fig. \ref{fig:residuals_15}. Here, we show a close-up of the residuals after all fits are subtracted (so not only \ce{CO2}, like in the left panel) and we plot three \ce{CO^18O} models at 200, 300 and 400 K, scaled with $R$ to fit the feature. We assume a column density of $N = 5\times10^{16}$ cm$^{-2}$ for these models, which is the expected column density assuming $N$(\ce{^13CO2})/$N$(\ce{CO^18O}) $\sim$ {4} (from the ISM abundance ratios of \ce{^12C}/\ce{^13C} $\sim 68$ and \ce{^16O}/\ce{^18O} $\sim 500${, accounting for the fact that \ce{CO2} has 2 O atoms}; \citealt{wilson1994, wilson1999, milam2005}). The coldest model fits the feature very nicely. This would be in line with the cold ($\sim$200 K) temperature found for the \ce{^13CO2} feature. However, we do wish to stress that this residual, as it is depicted in the right panel, could also contain some high-energy \ce{H2O} emission that is lacking from our best-fit model, as the comparison with DR Tau and DF Tau suggests. This could bias the determination of the temperature. Still, the prominence of the 15.07 $\mu$m feature suggests it should not be attributed solely to \ce{H2O} emission, and Fig. \ref{fig:residuals_15} demonstrates that a contribution from \ce{CO^18O} is certainly possible.

\FloatBarrier

\section{{Other emission features}} \label{app:ext}

\begin{figure*}
    \centering
    \includegraphics[width=\textwidth]{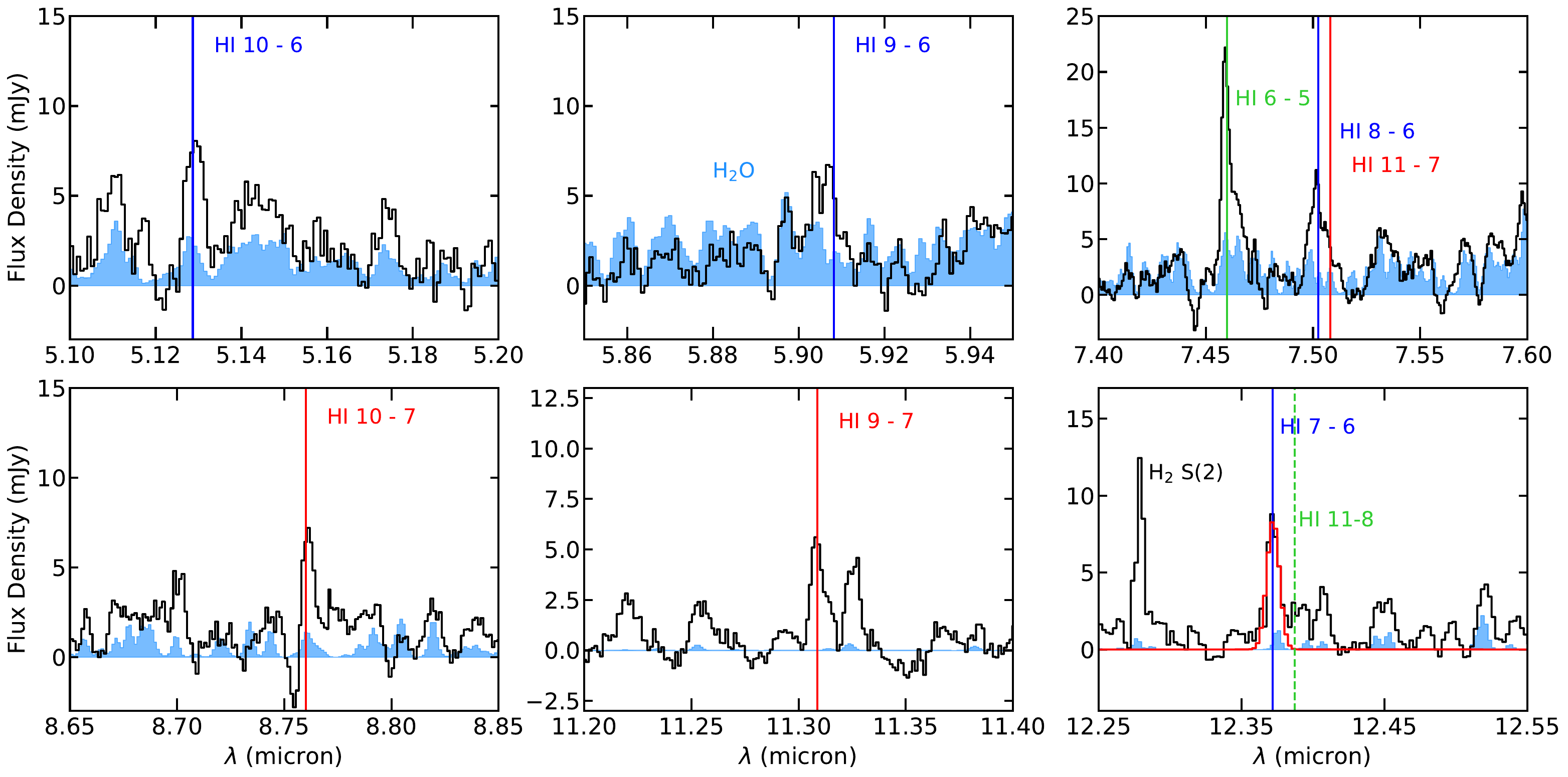}
    \caption{Zoom-ins of the CX Tau spectrum (black) around various possible detections of H lines. {We show the best-fit ro-vibrational \ce{H2O} model (the best fit to the 5.5-8.5 $\mu$m region) in all panels containing lines below 10 $\mu$m and the best-fit rotational model (the best fit to the 13.5-17.5 $\mu$m region) in all panels containing lines at longer wavelengths. These \ce{H2O} slab models are shown in blue}, and solid vertical lines indicate the {central} wavelengths of the H lines, where the $n_{\rm low} = 5$ (Paschen) series is shown in green, the $n_{\rm low} = 6$ (Humphreys) series is shown in blue and the $n_{\rm low} = 7$ series is shown in red.}
    \label{fig:H_lines}
\end{figure*}

\begin{figure*}
    \centering
    \includegraphics[width=0.9\textwidth]{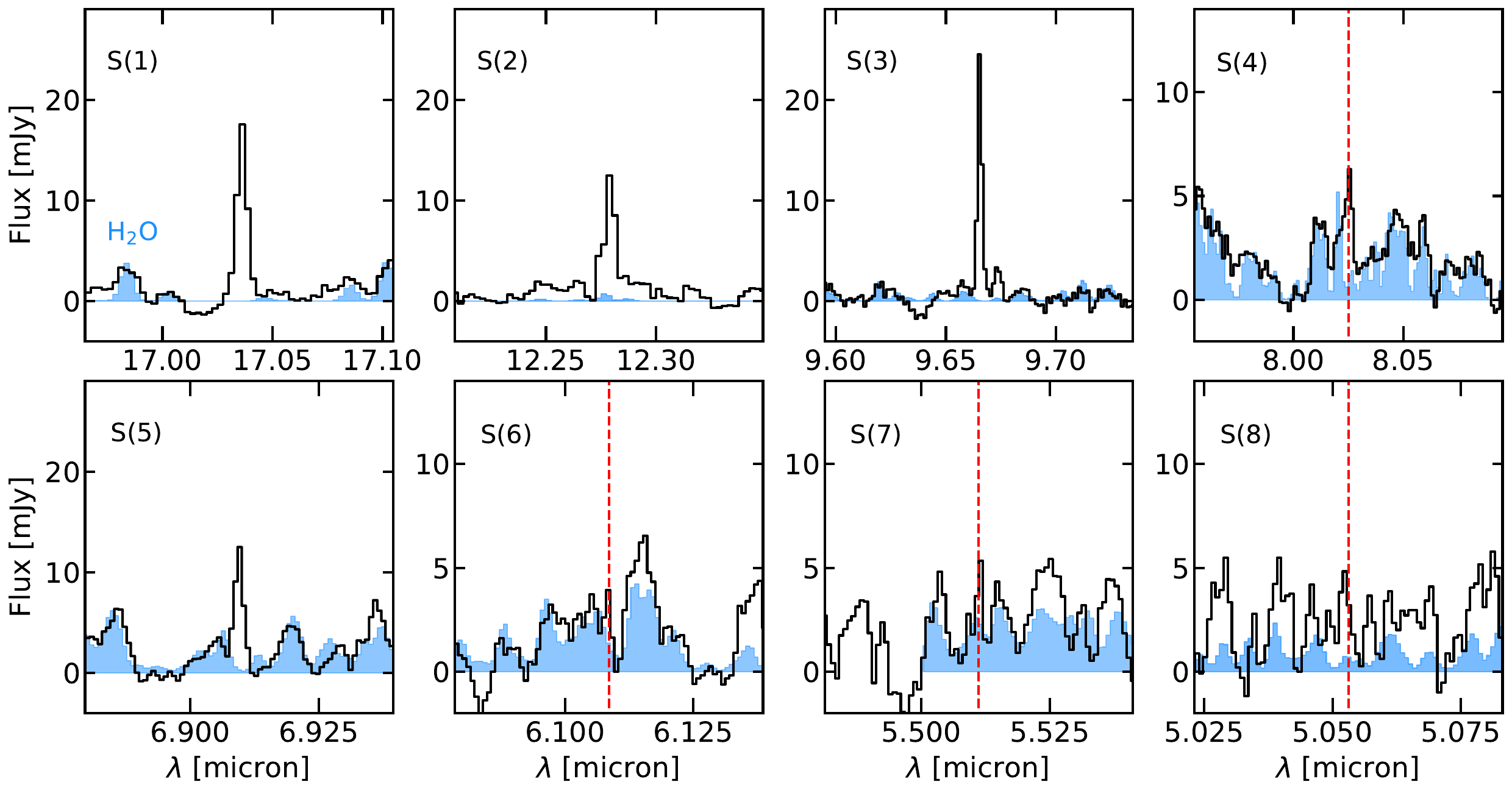}
    \caption{Zoom-ins of the CX Tau spectrum (black) at the locations of the 8 pure-rotational \ce{H2} lines in the MIRI/MRS wavelength range. {We show the best-fit ro-vibrational \ce{H2O} model (the best fit to the 5.5-8.5 $\mu$m region) in all panels containing lines below 10 $\mu$m and the best-fit rotational model (the best fit to the 13.5-17.5 $\mu$m region) in all panels containing lines at longer wavelengths. These \ce{H2O} slab models are shown in blue}.}
    \label{fig:H2_lines}
\end{figure*}

\begin{figure*}
    \centering
    \includegraphics[width=0.9\textwidth]{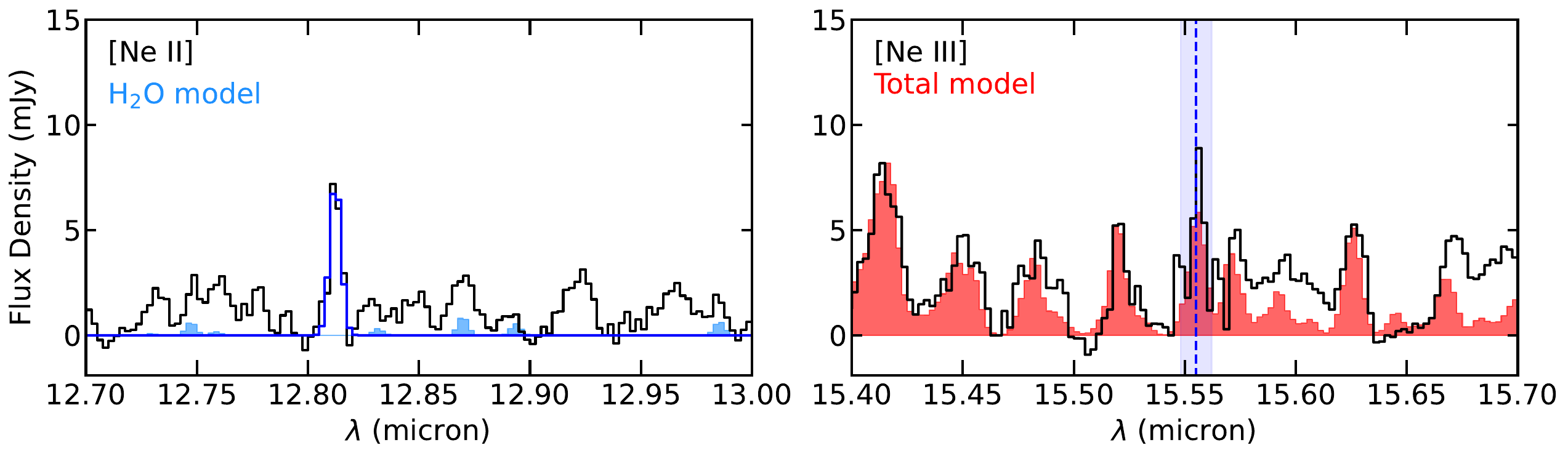}
    \caption{Zoom-ins of the CX Tau spectrum (black) around the [Ne II] and [Ne III] lines. In the left panel, a Gaussian fit to the [Ne II] line is shown in dark blue {together with the best-fit \ce{H2O} model from the 13.5-17.5 $\mu$m range in light blue}. In the right panel, the combined model containing \ce{^12CO2, ^13CO2, HCN, C2H2, H2O} and OH is shown in red. A blue dashed line indicates the position of the [Ne III] line and the shaded region surrounding it indicates the region in which the upper limit on its flux is derived.}
    \label{fig:Ne_lines}
\end{figure*}

\subsection{HI}\label{subsub:H}
We detect several atomic hydrogen recombination lines that have a principal quantum number of the lower energy levels of 5 or 6 (Pfund and Humphreys series), and a {tentatively detect a few} from the next higher series. An overview is presented in Fig. \ref{fig:H_lines}. The brightest H line detected is the Pf$\alpha$ line at 7.46 $\mu$m, but the Hu$\alpha$ (HI 7-6) line at 12.37 $\mu$m is also detected. The flux of this line has been shown by \citet{rigliaco2015} to have an empirical correlation with a source's accretion luminosity $L_{\rm acc}$, which is as follows: 
\begin{align}
    \log L_{\rm HI (7-6)}/L_\odot = (0.48 \pm 0.09) \times \log L_{\rm acc}/L_\odot - (4.68 \pm 0.10).
\end{align}
Then, following \citet{gullbring1998}, the derived $L_{\rm acc}$ can be used to calculate the mass accretion rate $\dot{M}_{\rm acc}$:
\begin{align}
    \dot{M}_{\rm acc} = \frac{L_{\rm acc}R_*}{GM_*}\left( 1 - \frac{R_*}{R_{\rm in}} \right)^{-1}.
\end{align}
Here, $M_* = 0.37 M_\odot$ is the stellar mass \citep{simon2017}, $R_{\rm in} = 5 R_*$ \citep{herczeg2008} is the radius at which the stellar magnetosphere truncates the accreting gas, and $R_* = 1.2 R_\odot$ is the stellar radius. The stellar radius is calculated using the Stefan-Boltzmann relation $L_* = 4\pi R_*^2 \sigma_{\rm B} T_{\rm eff}^4$, where the stellar luminosity $L_* = 0.2 L_\odot$ and the effective temperature is $T_{\rm eff} = 3483$ K \citep{herczeg2014}.\\
\newline
We measure the flux of the HI (7-6) line by fitting a Gaussian to it. A similar analysis was performed in \citet{franceschi2024}, where this line was found to be blended with the HI (11-8) line. We annotate the position of this line in Fig. \ref{fig:H_lines} to show that it is not detected in our data, hence we do not expect contribution from this line to impact our results. Calculating the accretion luminosity yields $L_{\rm acc} = (1.9\pm1.3)\times10^{-3} L_\odot$. Using this value to then calculate the mass accretion rate yields $\dot{M}_{\rm acc} = (2.6\pm1.9)\times10^{-10} M_\odot\; {\rm yr}^{-1}$. Both of these values are in good agreement with literature values \citep[within a factor 2-3, see, e.g.,][]{hartmann1998, fang2018}.

\subsection{\ce{H2}} \label{subsub:H2}
{Several pure-rotational \ce{H2} lines are detected in CX Tau. As shown in Fig. \ref{fig:H2_lines}, the 0-0 S(1), S(2), S(3), and S(5) lines are firmly detected. We find evidence that the emission from these lines is extended. This is demonstrated in \citet{anderson2024}, who show that this emission extends out to $\sim$500 au, much further than the CO gas disk detected with ALMA \citep{facchini2019}. It is possible that some of the \ce{H2} emission detected on source traces the disk itself, as \ce{H2} should naturally be abundant there. Additionally, the shape of the silicate feature indicates that the dust is more evolved and therefore perhaps less opaque, potentially aiding in such a detection, as was found to be the case in \citet{franceschi2024}. However, \ce{H2} emission is also a very common tracer of disk winds. We also detect [Ne II] in our source, which \citet{anderson2024} also demonstrate to be extended. Additionally, the work presents a detection of blue-shifted CO absorption with Keck/NIRSPEC. All of this indicates that CX Tau likely has a photo-evaporative wind.}

\subsection{Neon}\label{subsub:Ne}
Aside from hydrogen, emission from neon is also detected in CX Tau. As presented in Fig. \ref{fig:Ne_lines}, we clearly detect the [Ne II] line at 12.8 $\mu$m, and close inspection of the 15.5 $\mu$m region does reveal some emission that could potentially be associated with the [Ne III] line, but could also be slightly underproduced \ce{H2O} emission (combined contributions from \ce{^12CO2, ^13CO2, H2O, OH, C2H2}, and HCN are shown in red). We obtain the line flux of the [Ne II] line by fitting a Gaussian profile and we derive a $3\,\sigma$ upper limit on the [Ne III] line flux, after which we divide the two to obtain a $3\,\sigma$ upper limit on the [Ne III]-to-[Ne II] ratio. This yields a value of 0.5. \\
\newline
Theoretical work has shown that this value can indicate whether X-rays or extreme ultraviolet (EUV) radiation produces the emission. \citet{hollenback2009} showed that EUV radiation produces [Ne III]-to-[Ne II] ratios > 1, whereas \citet{glassgold2007} and \citet{meijerink2008} showed that X-rays produce [Ne III]-to-[Ne II] ratios of approximately 0.1. This difference is mainly caused by the difference in depth to which X-ray or EUV photons can penetrate the disk. X-ray photons penetrate further, into a layer that contains mostly neutral H. This leads to a very effective charge exchange between Ne$^{2+}$ and H converting a lot of the Ne$^{2+}$ into Ne$^+$, thus leading to a low [Ne III]-to-[Ne II] ratio. EUV photons, however, cannot penetrate that deeply into the disk, only reaching a layer that is still mostly ionized, containing much less neutral H. As such, the charge exchange converting Ne$^{2+}$ into Ne$^+$ is much less efficient, leading to much higher [Ne III]-to-[Ne II] ratios, typically higher than 1. \\
\newline
Our derived upper limit on the [Ne III]-to-[Ne II] ratio is less than 1, indicating that the former case, where the neon ions are predominately produced by X-rays, is more likely to be responsible for the observed Ne emission. This is in line with typical findings for T Tauri stars, such as previous results from \textit{Spitzer} \citep{szulagyi2012, espaillat2013} and more recent JWST results \citep{espaillat2023}. {However, \citet{anderson2024} point out that the relative line fluxes of the \ce{H2} S(1) and S(3) lines are more consistent with EUV excitation, rather than X-rays. Our derived [Ne III]-to-[Ne II] ratio (while more consistent with X-ray excitation) could indicate that CX Tau is an in-between case, making it an interesting subject for further study.} We do not detect any other atomic emission from, e.g., argon or iron.\\

\end{appendix}

\end{document}